\documentclass[aps,reprint,twocolumn,superscriptaddress,citeautoscript,showpacs,amsmath,amssymb,floatfix,prb,longbibliography]{revtex4-1}
\usepackage{}%\bibliographystyle{apsrev}
\usepackage{graphicx}
\usepackage{amsmath}
\usepackage{amssymb}
\usepackage{mathtools}
\usepackage{textcomp,gensymb} %For degree symbol
\usepackage{bm} %For bold math (Greek) symbols or bold italic in math mode
\usepackage[usenames,dvipsnames]{xcolor}
\usepackage{float}
\usepackage{setspace}
\usepackage{dcolumn}
\usepackage{array}

\usepackage{changepage}
\usepackage{tikz}
\usepackage{contour}
\usepackage[left]{lineno}
\usepackage{blindtext}
\usepackage{setspace}
\usepackage[none]{hyphenat}%for removing hyphen
\usepackage[colorlinks=true,citecolor=blue,linkcolor=blue,pdftex,hypertexnames=false,urlcolor=blue,linktocpage=true]{hyperref}%for hyperlink
\definecolor{ashgray}{rgb}{0.7,0.75,0.71}
\definecolor{mspringgreen}{rgb}{0, 0.8, 0.1}
\definecolor{auburn}{rgb}{0.43, 0.21, 0.1}
\definecolor{ao(english)}{rgb}{0.0, 0.5, 0.0}
\definecolor{afw}{rgb}{0.95, 0.95, 0.96}
\definecolor{magnolia}{rgb}{0.97, 0.96, 1.0}
\definecolor{wsmk}{rgb}{0.96, 0.96, 0.96}

\newcolumntype{M}[1]{>{\centering\arraybackslash}m{#1}}

\usepackage{dcolumn}% Align table columns on decimal point
\usepackage{multirow} 
\usepackage{soul}
\usepackage{titlesec}
\usepackage{color}

\usepackage{bbold}

\newcolumntype{d}[1]{D{.}{.}{#1}}

\begin{document}

\title{Suppression of the valence transition in solution-grown single crystals of Eu$_2$Pt$_6$Al$_{15}$}

\author{Juan Schmidt}
\affiliation{Department of Physics and Astronomy, Iowa State University, Ames, IA 50011, USA}
\affiliation{Ames National Laboratory, Iowa State University, Ames, IA 50011, USA}
\author{Dominic H. Ryan}
\affiliation{Department of Physics, McGill University, 3600 University Street, Montreal Quebec Canada H3A 2T8}
\author{Oliver Janka}
\affiliation{Universität des Saarlandes, Anorganische Festkörperchemie, Campus C4 1, 66123 Saarbrücken, Germany}
\author{Jutta Kösters}
\affiliation{Institut für Anorganische und Analytische Chemie, Universität Münster, Corrensstrasse 20, 48149 Münster, Germany}
\author{Carsyn L. Mueller}
\affiliation{Department of Physics and Astronomy, Iowa State University, Ames, IA 50011, USA}
\affiliation{Ames National Laboratory, Iowa State University, Ames, IA 50011, USA}
\author{Aashish Sapkota}
\affiliation{Ames National Laboratory, Iowa State University, Ames, IA 50011, USA}
\author{Rafaela F. S. Penacchio}
\affiliation{Department of Physics and Astronomy, Iowa State University, Ames, IA 50011, USA}
\affiliation{Ames National Laboratory, Iowa State University, Ames, IA 50011, USA}
\affiliation{Institute of Physics, University of Sao Paulo, Sao Paulo, SP, Brazil}
\author{Tyler J. Slade}
\affiliation{Ames National Laboratory, Iowa State University, Ames, IA 50011, USA}
\author{Sergey L. Bud'ko}
\affiliation{Department of Physics and Astronomy, Iowa State University, Ames, IA 50011, USA}
\affiliation{Ames National Laboratory, Iowa State University, Ames, IA 50011, USA}
\author{Paul C. Canfield}
\affiliation{Department of Physics and Astronomy, Iowa State University, Ames, IA 50011, USA}
\affiliation{Ames National Laboratory, Iowa State University, Ames, IA 50011, USA}

\date{\today}

\pacs{1234}

\begin{abstract}

The study of Eu intermetallic compounds has allowed the exploration of valence fluctuations and
transitions in 4f electron systems. Recently, a Eu$_2$Pt$_6$Al$_{15}$ phase synthesized by arc-melting followed
by a thermal treatment was reported [M. Radzieowski \textit{et al.}, 
J Am Chem Soc 140(28), 8950-8957 (2018)], which undergoes a transition upon cooling below 45~K that was interpreted as a valence transition from Eu$^{2+}$ to Eu$^{3+}$. In this paper, we present the discovery of another polymorph of Eu$_2$Pt$_6$Al$_{15}$ obtained by high temperature solution growth, which presents different physical properties than the arc-melted polycrystalline sample. Despite the similarities in crystal structure and chemical composition, the Eu valence transition is almost fully suppressed in the solution-grown crystals, allowing the moments associated with the Eu$^{2+}$ state to order antiferromagnetically at around 14~K. A detailed analysis of the crystal structure using single crystal X-ray diffraction reveals that, although the solution grown crystals are built from the same constituent layers as the arc-melted samples, these layers present a different stacking. The effect of different thermal treatments is also studied. Different anneal procedures did not result in significant changes in the intrinsic properties, and only by arc-melting and quenching the crystals we were able to convert them into the previously reported polymorph.

\end{abstract}

\maketitle
\section{Introduction} 
\label{sec:Introduction}

Among rare-earth compounds, those containing Ce, Sm, Eu or Yb are of particular interest because they commonly exhibit different valences and intriguing related phenomena. The simplest case is inhomogeneous mixed valence, observed in cases such as Eu$_3$O$_4$ \cite{Rau1966}, Eu$_3$S$_4$ \cite{Carter1972,Batlogg1976}, Sm$_3$S$_4$ \cite{Carter1972,Batlogg1976,Batlogg1976b}, and EuPtP \cite{Lossau1989}, where two different crystallographic sites present different static valences. Other compounds may exhibit valence fluctuations, leading to rare-earth ions on a given site with intermediate mean valence, as is the case for $R$Cu$_2$Si$_2$ ($R=$Ce, Eu, Yb) \cite{Sales1976}, EuRh$_2$P$_2$ \cite{Michels1996}, $R$Ni$_2$Ge$_2$ ($R=$Ce, Yb) \cite{Budko1999} or Yb$_9$Pt$_4$Ga$_{24}$ \cite{Sichevych2017}. 

\begin{figure}[H]
 \centering
 \includegraphics[width=\linewidth]{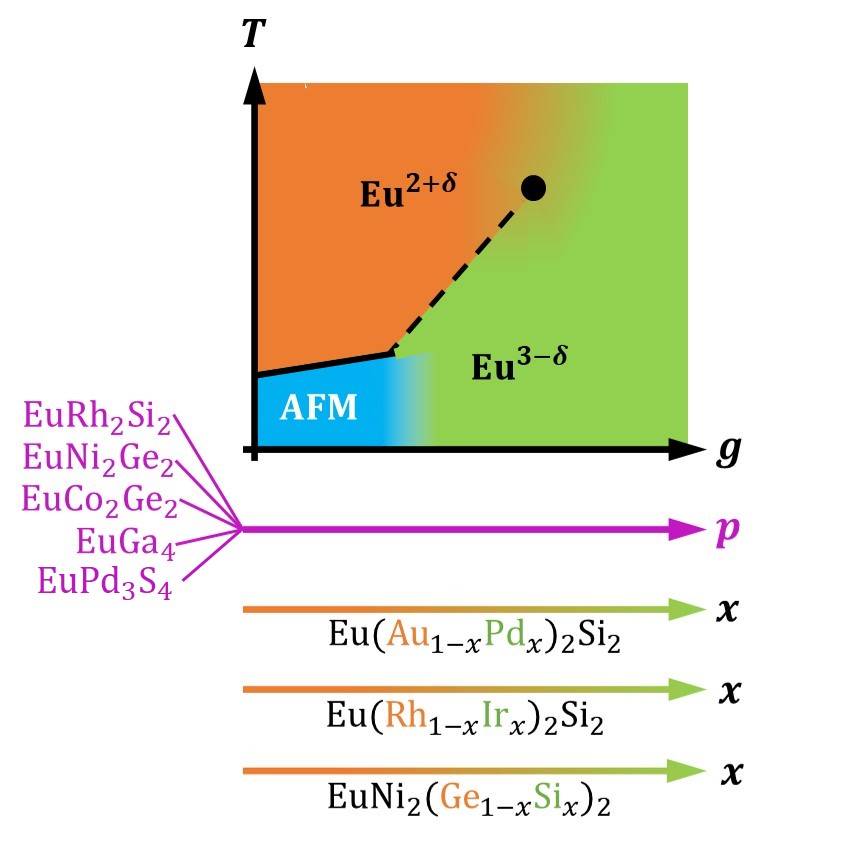}
 \caption{{Schematic phase diagram for different Eu compounds as a function of temperature, $T$, and a non-thermal tuning parameter, $g$, such as pressure \cite{Honda2016,Honda2018,Huyan2023} or chemical composition \cite{Segre1982,Seiro2011,Wada1999}. The solid black line represents the second-order antiferromagnetic transition, the dashed black line represents the first order the valence transition, which can reach a critical endpoint, represented by the solid black circle, beyond which the valence change becomes continuous. }}
 \label{fig:phase_diagram}
\end{figure}

Some systems with these rare-earth elements may change their valence with temperature or pressure, either continuously, as EuPtAl \cite{Lossau1989} and Eu$_4$Bi$_6$Se$_{13}$ \cite{Xu2025}, or undergo valence transitions, such as EuPd$_3$S$_4$ \cite{Huyan2023} and EuSn$_2$As$_2$ \cite{Sun2021}. Many of those with ThCr$_2$Si$_2$ structure present a common general behavior upon changes in pressure or changes in chemical composition. In the compounds Eu$M_2X_2$ ($M=$ Rh, Ni, Co; $X=$ Si, Ge) at ambient pressure, Eu ions are nearly divalent and order antiferromagnetically at a temperature $T_N$; however, under pressure, a first-order valence transition emerges at a higher temperature, below which the nearly trivalent (non-moment-bearing) Eu ions do not order any more \cite{Honda2016,Honda2018}. A similar pattern occurs with the variation of chemical composition in systems like Eu(Pd$_{1-x}$Au$_{x}$)$_2$Si$_2$ \cite{Segre1982} and Eu(Rh$_{1-x}$Ir$_x$)$_2$Si$_2$ \cite{Seiro2011}. The schematic phase diagram shown in Fig. \ref{fig:phase_diagram} summarizes the general behavior of many systems which can be tuned from having antiferromagnetic ordering to valence transitions to continuous valence changes by controlling a nonthermal parameter $g$, such as pressure or chemical composition.

Recently, Eu$_2$Pt$_6$Al$_{15}$ was reported to exhibit a valence transition from Eu$^{2+}$ to Eu$^{3+}$ upon cooling below $\sim$50~K \cite{Radzieowski2018}, based on M\"ossbauer spectroscopy, resistance and magnetic susceptibility measurements. This phase was obtained in polycrystalline form by arc-melting Eu, Pt and Al together, followed by a thermal treatment. The rest of the members of the $R_2$Pt$_6$Al$_{15}$ and $R_2$Pt$_6$Ga$_{15}$ series ($R=$Sc, Y, La-Lu) were also studied in a polycrystalline form \cite{Oyama2020,Radzieowski2020}, and it was shown that at least Sc$_2$Pt$_6$Al$_{15}$ and Ho$_2$Pt$_6$Al$_{15}$ crystallize in the same structure as Eu$_2$Pt$_6$Al$_{15}$. This is a superstructure of the Sc$_{0.6}$Fe$_2$Si$_{4.9}$-type structure (hexagonal $P6_3/mmc$) \cite{Kotur1991,Latturner2002,Niermann2004}, where the order of the vacancies can be interpreted as a (3+1)D modulation belonging to the orthorhombic superspace group $Cmcm(\alpha,0,0)0s0$ ($\alpha=2/3a^*$), with a monoclinic approximant with space group $P2_1/m$ \cite{Radzieowski2017}. No single crystal synthesis or characterization was reported for any of these compounds, except for solution-grown Ce$_2$Pt$_6$Al$_{15}$ crystals that exhibit physical properties different from those of polycrystalline arc-melted Ce$_2$Pt$_6$Al$_{15}$ \cite{Ota2023}.

The effect of chemical substitution on the valence transition of Eu$_2$Pt$_6$Al$_{15}$ has also been studied. For Eu$_2$Pt$_6$(Al$_{1-x}$Ga$_x$)$_{15}$, the valence transition is gradually suppressed with increasing $x$, eventually allowing the divalent Eu ions to order antiferromagnetically at $T_N\sim 15$~K \cite{Oyama2020}. The corresponding phase diagram is consistent with the one schematized in Fig. \ref{fig:phase_diagram} as well. Partial substitution of Pt with Pd slightly suppresses the valence transition as well, whereas substitution with Ir slightly enhances it \cite{Engel2024}. 

In this work, we present a way of obtaining single-crystalline Eu$_2$Pt$_6$Al$_{15}$ by the high-temperature solution growth method. We show that the chemical composition of these solution-grown crystals is the same as the previously reported phase obtained by arc-melting \cite{Radzieowski2018}, and that no substantial differences in the crystal structure can be discerned by powder X-ray diffraction. However, the magnetic and electronic properties of the solution-grown crystals are drastically different: The valence transition is suppressed, allowing the magnetic Eu ions to order antiferromagnetically. A more detailed study of the crystal structure using single crystal X-ray diffraction shows that, although both phases correspond to superstructures of the Eu$_{0.67}$Pt$_2$Al$_5$ average structure and are built of the same layers, these layers are stacked in a different way. The way these structural differences impact the rest of the physical properties is discussed and compared to the other related systems mentioned.

\section{Experimental details}
\label{sec:Experimental}

Single crystals of Eu$_2$Pt$_6$Al$_{15}$ were obtained by the high-temperature solution growth method \cite{Canfield2020} out of a ternary melt with excess of Al. The starting materials were elemental Pt powder (Ames National Laboratory, 99.9+\% purity), pieces of Al rod (Alfa Aesar, 99.999\% purity) and Eu pieces (Materials Preparation Center - Ames National Laboratory, 99.99+\% purity), combined in the ratio Eu$_5$Pt$_{10}$Al$_{85}$.

\begin{figure}[H]
 \centering
 \includegraphics[width=\linewidth]{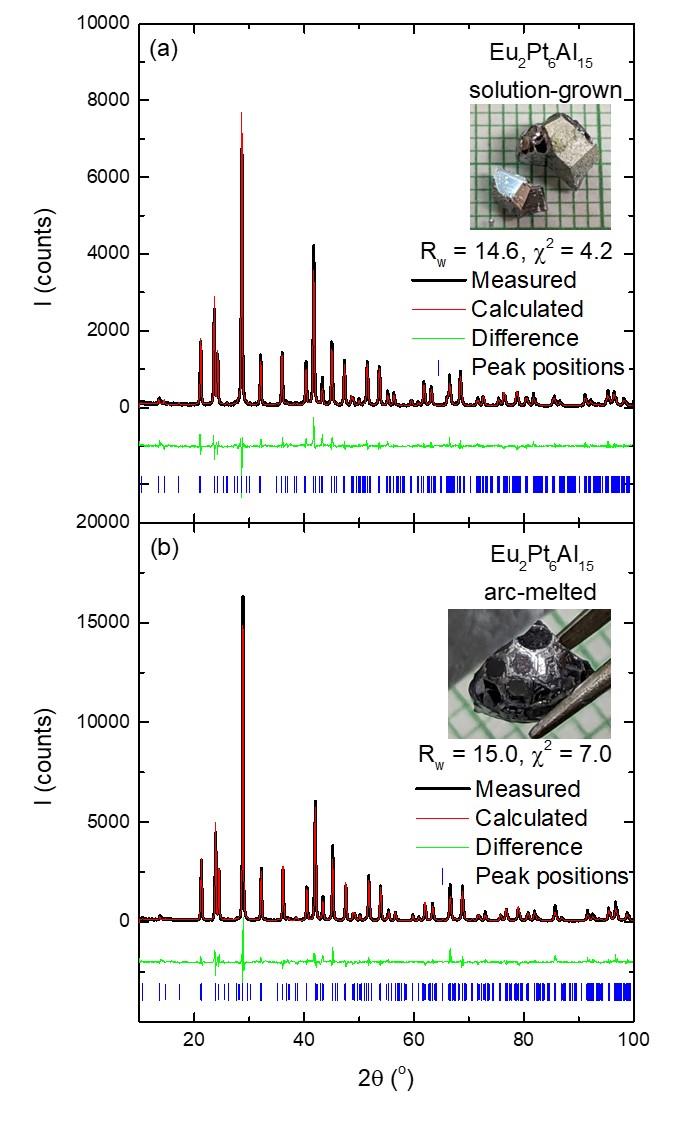}
 \caption[Powder XRD of Eu$_{2}$Pt$_{6}$Al$_{15}$]{{(a) Powder X-ray diffraction pattern of the Eu$_{2}$Pt$_{6}$Al$_{15}$ solution-grown phase (black), the best fit obtained by Rietveld refinement (red), the residues (green), and the peak positions (blue). A photograph of two crystals is also shown. (b) Powder X-ray diffraction pattern of the Eu$_{2}$Pt$_{6}$Al$_{15}$ phase obtained after arc-melting the SG crystals (black), the best fit obtained by Rietveld refinement (red), the residues (green), and the peak positions (blue). A photograph a piece of the arc-melted button is also shown; note the faceting on the arc-melted piece.}}
 \label{fig:pdxrd_hex}
\end{figure}

There is a large exothermic reaction between Al and Pt/Pd. To prevent this exothermic reaction from occurring inside the alumina crucible in which the crystals grow, and potentially cracking the former, a similar step was followed to that reported for the growth of the Al-Pd-Mn single grain quasicrystal \cite{Fisher1999}. This consisted of pre-alloying the Al and Pt using an arc furnace. Pt powder was first pressed into a pellet, arc-melted, weighed and then arc-melted in contact with Al pieces, to form a single button with the proportion 10:85 of Pt:Al. The arc-melted button was placed in a 2 ml alumina fritted Canfield Crucible Set \cite{CanfieldP.C.KongT.KaluarachchiU.S.2016,LSPCeramics} together with the corresponding amount of Eu to achieve the desired initial composition. The crucible set was sealed in a fused silica tube with a partial pressure of 1/6 atm of Ar and placed inside a box furnace. It was heated to 1180 $^{\circ}$C, kept at that temperature for 5 hours, and slowly cooled down to 900 $^{\circ}$C over the course of 60 hours. At 900 $^{\circ}$C the flux was decanted with the aid of a centrifuge. A picture of the obtained crystals are shown in Fig. \ref{fig:pdxrd_hex}(a).

Given that we found the properties of the solution-grown (SG) Eu$_2$Pt$_6$Al$_{15}$ to be significantly different from those of the reported arc-melted (AM) polycrystalline samples, we synthesized polycrystalline Eu$_2$Pt$_6$Al$_{15}$ samples by arc-melting approximately 500 mg of SG Eu$_2$Pt$_6$Al$_{15}$ single crystals. Our arc-melted sample formed surface facets upon cooling, as can be seen in the inset of Fig. \ref{fig:pdxrd_hex}(b).

In addition, the effects of different thermal treatments were studied on both the SG and AM Eu$_2$Pt$_6$Al$_{15}$. The SG crystals were placed in a 2 ml alumina crucible and sealed in a fused silica tube with 1/3 atm of Ar, and annealed in a box furnace for 86 hours at 600 $^{\circ}$C, after which the furnace was turned off and the ampoules were allowed to cool slowly inside. A second anneal step was attempted for the same time at 750 $^{\circ}$C, resulting in the decomposition of the crystal surface (most likely due to the melting of small droplets of Al-rich flux on the crystal surfaces for temperatures above 660 $^{\circ}$C). To avoid this, a third thermal treatment was attempted, referred to as \textit{in situ anneal}. This consisted of repeating the procedure to grow the single crystals, but adding an additional 100-hour dwell at the final temperature (900 $^{\circ}$C) before decanting the flux. Finally, the AM polycrystalline samples were also annealed at 600 $^{\circ}$C following the same procedure as the first anneal described here.

The concentration levels of each element were determined by energy dispersive X-ray spectroscopy (EDS) quantitative chemical analysis using an EDS detector (Thermo NORAN Microanalysis System, model C10001) attached to a JEOL scanning-electron microscope (SEM). An acceleration voltage of
$16\ \text{kV}$, working distance of $10\ \text{mm}$ and take-off angle of $35 ^{\circ}$ were used to measure all standards and crystals with unknown composition. A single crystal of EuAl$_4$ \cite{Ryan2024} was used as a standard for Eu and Al quantification, and a single crystal of PtTe$_2$ was used as a standard for Pt quantification. The spectra were fitted using NIST-DTSA II Microscopium software \cite{Newbury2014}. The composition of each crystal was measured at four different positions on an arbitrary crystal's face. The average compositions and error bars were obtained from these data, corresponding to the 95\% confidence interval accounting for both inhomogeneity and goodness of fit of each spectra. 

Powder X-ray diffraction (PXRD) measurements were performed using a Rigaku MiniFlex II powder diffractometer with Cu K$\alpha$ radiation ($\lambda=1.5406\ \text{\AA}$). For each growth (SG, AM, etc.), a few crystals were finely ground to a powder and dispersed evenly on a single-crystalline Si zero background holder, with the aid of a small quantity of vacuum grease. Intensities were collected for $2\theta$ ranging from $10^{\circ}$ to $100^{\circ}$, in step sizes of $0.01^{\circ}$, counting for 4 seconds at each angle. Rietveld refinement was performed on each diffractogram using the GSAS II software package \cite{Dreele2014}. Refined parameters included but were not limited to lattice parameters, atomic positions, isotropic displacement parameters and site occupancies.

For single-crystal X-ray diffraction (SCXRD) measurements, regularly shaped crystal fragments of Eu$_2$Pt$_6$Al$_{15}$ were obtained by fracturing the solution-grown crystals. The fragments were attached to quartz fibers using beeswax. The data sets were collected on a Stoe IPDS-II diffractometer (graphite monochromatized Mo $K\alpha$ radiation; oscilation mode). Numerical absorption correction \cite{Stoe2014} was applied to the data series. Other fragments of Eu$_2$Pt$_6$Al$_{15}$ were also studied using a
Rigaku XtaLab Synergy-S diffractometer with Ag radiation ($\lambda=0.56087\ \text{\AA}$), in transmission mode, operating at 65~kV and 0.67 mA. The samples were held on a nylon loop with Apiezon N grease. The total number of runs and images was based on the strategy calculation from the program CrysAlisPro (Rigaku OD, 2023).

DC magnetization measurements were carried out on a Quantum Design Magnetic Property Measurement System (MPMS classic and MPMS3) superconducting quantum interference device (SQUID) magnetometer (operated in the range $1.8\ \text{K}\leq T \leq 300\ \text{K}$, $|H|\leq 50\ \text{kOe}$). Each sample was measured with the field applied in different directions relative to the crystallographic axes, which were determined with a Laue camera. Most measurements were performed under zero-field cooling (ZFC) protocols, by setting the field to zero at 60~K, after having centered the sample, and subsequently cooling down to the lowest temperature involved in the measurement, then applying the field, and finally measuring as a function of temperature or field. Field cooling (FC) protocols were also employed in some cases to verify that their behavior coincided with that measured with the ZFC protocol. The samples were glued to a Kel-F disk which was placed inside a plastic straw; the contribution of the disk to the measured magnetic moment was independently measured in order to subtract it from our results. 

 The temperature-dependent AC resistance of the samples was measured using a Quantum Design Physical Property Measurement System (PPMS) using the AC transport (ACT) option, with a frequency of $17\ \text{Hz}$ and a $3\ \text{mA}$ excitation current. The resistance both parallel and perpendicular to the highest symmetry direction (labeled $c$ for all the systems studied) was measured using a standard four-contact geometry. Electrical contacts with less than $1.5\ \Omega$ resistance were achieved by spot welding $25\ \mu\text{m}$ Pt wire to the samples, followed by adding Epotek H20E silver epoxy, and curing the latter for 1 hour at $120^{\circ}\text{C}$.

$^{151}$Eu M\"ossbauer spectroscopy measurements were carried out using a 4 GBq $^{151}$SmF$_3$ source, driven in sinusoidal mode.
  source, driven in sinusoidal mode.  The drive motion was calibrated using a
standard $^{57}$Co\underline{Rh}/$\alpha$-Fe foil. Isomer shifts are referenced
relative to EuF$_3$ at ambient temperature. A thin (0.25~inch) NaI(Tl)
scintillation detector was employed to detect the transmitted gamma rays.
$\sim$120~mg of Eu$_2$Pt$_6$Al$_{15}$ was hand-ground in an agate mortar under hexane to
protect from oxidation. The powder was mixed with boron nitride to make a uniform absorber
and loaded into a thin-window delrin holder. The sample was cooled in a
vibration-isolated closed-cycle helium refrigerator with the sample in helium
exchange gas. The spectra were fitted to a sum of Lorentzian lines with
the positions and intensities derived from a full solution to the nuclear
Hamiltonian \cite{voyer}. Typical line widths (HWHM) were 1.1–1.3 mm/s and
both valence forms showed a quadrupole interaction of about +5 mm/s. Both the
line widths and quadrupole interactions were essentially temperature independent.

\section{Results and discussions}
\label{sec:Results}

\subsection{Chemical and structural similarities}

A quantitative EDS analysis of the solution-grown (SG) crystals and arc-melted (AM) polycrystalline samples revealed that they have similar compositions and close to the expected atomic fractions for the stoichiometry Eu$_2$Pt$_6$Al$_{15}$. Table \ref{tab:EDS_hex} shows that the composition of the SG crystals agrees with the composition of the AM sample within the uncertainty given by the EDS quantitative analysis, as well as with the expected stoichiometric ratio given in the last column. These values are also consistent to the EDS results previously reported for the arc-melted sample \cite{Radzieowski2018}.

\begin{table}[H]
\centering
\caption{\label{tab:EDS_hex} Elemental analysis of Eu$_{2}$Pt$_{6}$Al$_{15}$ SG phase, in comparison with the Eu$_{2}$Pt$_{6}$Al$_{15}$ AM phase.}
\begin{tabular}{cccc}
\hline
\hline
\textrm{Element}&
\textrm{Solution grown}& \textrm{Arc-melted} \footnote{obtained by arc-melting solution grown crystals}& \textrm{Nominal}\\
\hline
Eu & 0.087(7) & 0.080(5) & 0.087 \\
Pt & 0.27(2) & 0.26(2) & 0.26 \\
Al & 0.64(6) & 0.65(7) & 0.65 \\
\hline
\hline
\end{tabular}
\end{table}

 monoclinic $P2_1/m$ approximant reported for Eu$_2$Pt$_6$Al$_{15}$ \cite{Radzieowski2018,Radzieowski2017} yield similar goodness-of-fits. No obvious structural differences can be appreciated in these powder XRD measurements, other than very subtle changes in the lattice parameters, shown in Tab. \ref{tab:latticeparam}. The high-symmetry axis was chosen to be along $c$ instead of $b$ for an easier comparison with the hexagonal averaged structure reported in previous works \cite{Latturner2002,Niermann2004,Radzieowski2018,Radzieowski2020}.

\begin{table}[H]
\centering
\caption{\label{tab:latticeparam} Lattice parameters obtained from the Rietveld refinement of the Eu$_{2}$Pt$_{6}$Al$_{15}$ SG and AM phase, using the monoclinic $P112_1/m$ approximant structural model.}
\begin{tabular}{ccc}
\hline \hline
\textrm{Element}&
\textrm{Solution grown}& \textrm{Arc-melted} \\
\hline
$a$ & 7.46(2) \AA & 7.444(8) \AA \\
$b$ & 7.46(2) \AA & 7.44(1) \AA \\
$c$ & 16.642(1) \AA & 16.664(2) \AA \\
$\gamma$ & 119.96(4)$^{\circ}$ & 119.94(2)$^{\circ}$\\
$V$ & 803.2(2) \AA$^3$ & 800.0(2) \AA$^3$\\
\hline \hline
\end{tabular}
\end{table}

\subsection{Electronic and magnetic differences}

\begin{figure}[H]
 \centering
 \includegraphics[width=\linewidth]{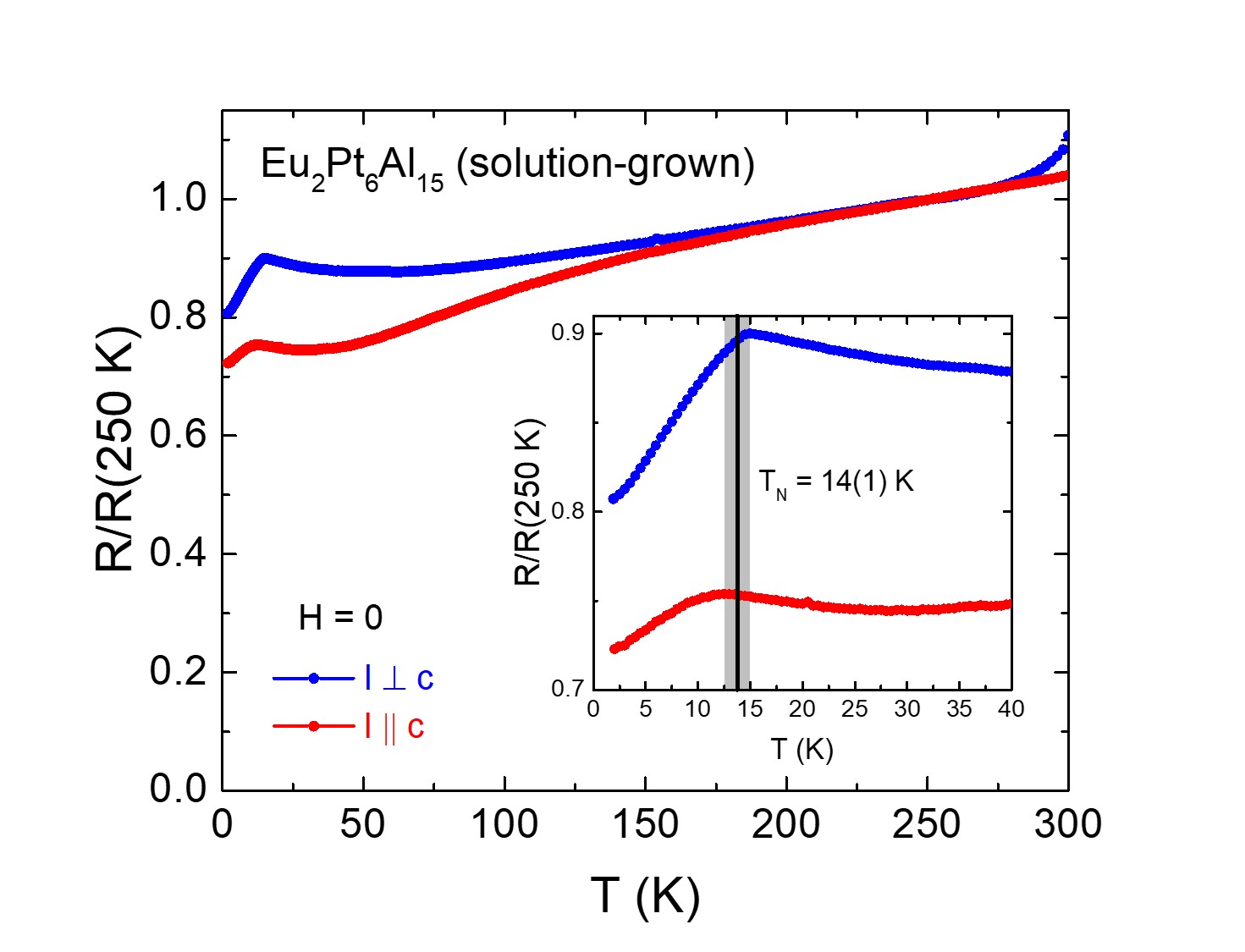}
 \caption{{Main panel: Resistance measured upon warming normalized to its value at 250~K of the Eu$_2$Pt$_6$Al$_{15}$ phase obtained by solution growth with the current applied perpendicular (blue) and parallel (red) to $c$. Inset: enlarged scale in order to indicate the position of the resistance maxima (vertical black line), corresponding to $T_N$. The uncertainty (indicated by the gray region) is determined by the difference between the maxima for both current directions.}}
 \label{fig:RT_hex}
\end{figure}

The AM sample was reported to have a valence transition at around 50~K that manifests itself as a step-like decrease in the resistivity due to the loss of moments that can cause additional scattering in their paramagnetic state \cite{Radzieowski2018}, since the low-temperature Eu ions lose their magnetic moment when they change their valence. However, the SG crystals display a kink-like feature at 14.7(5)~K, as seen in blue and red in the main panel of Fig. \ref{fig:RT_hex} for the current perpendicular and parallel to the $c$-axis, respectively. The temperature at which the resistance becomes locally maximum was taken as the transition temperature. Since this value was slightly different for the measurements with current perpendicular and parallel to $c$, the average was calculated, and the halved difference was taken as the uncertainty. These are indicated by the black line and the gray region in the inset of Fig. \ref{fig:RT_hex}, respectively.

\begin{figure}[H]
 \centering
 \includegraphics[width=\linewidth]{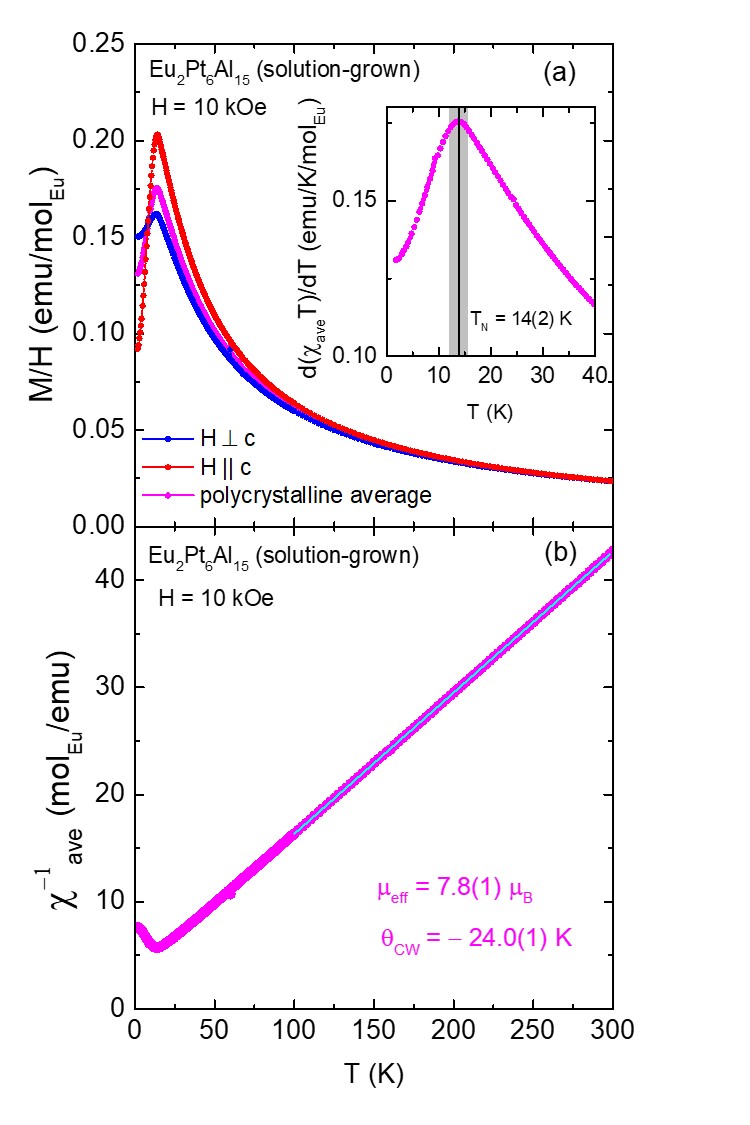}
 \caption{{(a) Temperature dependent ZFC and FC magnetization by the applied field of 10~kOe of the solution-grown Eu$_2$Pt$_6$Al$_{15}$ phase with the field applied perpendicular (blue) and parallel (red) to $c$, as well as for the polycrystalline average (magenta, see main text for details). The inset shows the polycrystalline average in an enlarged scale, with the transition temperature indicates with a black line and the uncertainty with a gray rectangle. (b) Polycrystalline average of the inverse susceptibility as a function of temperature for the solution-grown Eu$_2$Pt$_6$Al$_{15}$ phase (magenta); the linear fit is plotted with a cyan line.}}
 \label{fig:MT_hex}
\end{figure}

\begin{figure}[H]
 \centering
 \includegraphics[width=\linewidth]{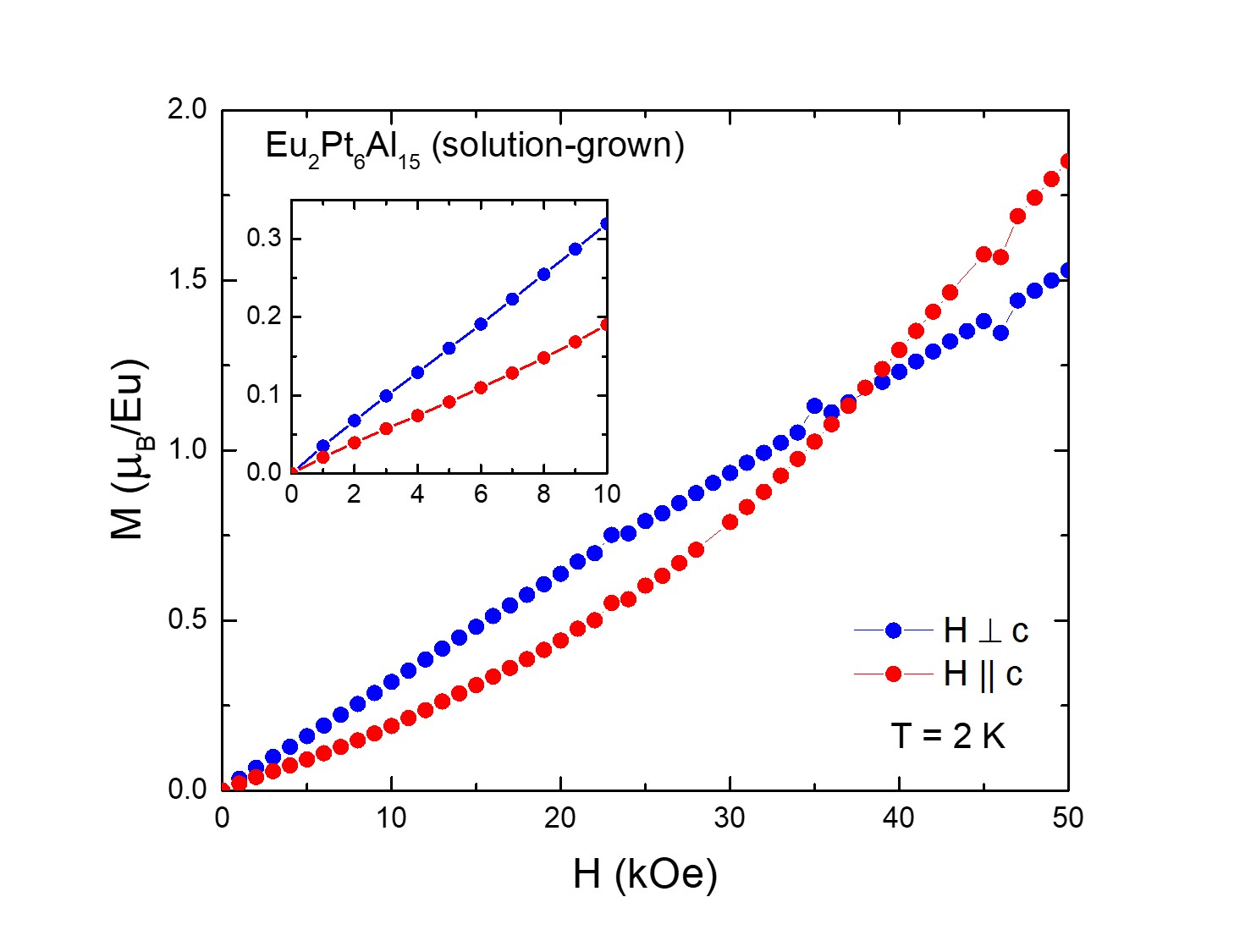}
 \caption{{Magnetization of Eu$_{2}$Pt$_{6}$Al$_{15}$ as a function of the applied field, oriented perpendicular (blue) and parallel (red) to $c$, measured at 2~K (main panel) and at 30~K (inset).}}
 \label{fig:MH_hex}
\end{figure}

In order to get a better understanding of the low temperature state of the solution-grown single crystals, temperature- and field-dependent magnetization measurements were performed and are shown in Figs. \ref{fig:MT_hex} and \ref{fig:MH_hex}, respectively. The main panel in Fig. \ref{fig:MT_hex}(a) shows the temperature-dependent magnetization measured on the solution-grown crystals with a field of 10~kOe applied perpendicular (blue) and parallel (red) to the $c$ axis. They were measured under ZFC and FC protocols which perfectly overlap with each other, both suggesting the presence of an antiferromagnetic phase transition at the temperature where the magnetization is maximum. The main panel in Fig. \ref{fig:MH_hex} shows the field-dependent magnetization of the low-temperature state, measured at 2~K, for the field applied perpendicular (blue) and parallel (red) to $c$. An enlarged view of the low-field data is plotted in the inset, showing that there is no remanent magnetization at zero field, further ruling out the possibility of having ferromagnetic/ferrimagnetic components in the low-temperature ordered state. Together, the $M(T)$ and $M(H)$ dependencies shown in Figs. \ref{fig:MT_hex}(a) and \ref{fig:MH_hex} indicate that the system undergoes an antiferromagnetic transition. Neither $\chi_a(T)$ nor $\chi_b(T)$ drop to zero for $T\rightarrow0$, which complicates the determination of the moment direction in the order state based solely on these results. The convexity of $M(H)$ for $H||c$ shown in red in the main panel of Fig. \ref{fig:MH_hex} suggests that there may be a metamagnetic transition at fields beyond 50~kOe.
%The possibility of having ferromagnetic/ferrimagnetic components in the low-temperature ordered state can be discarded since there the $M(H)$ measured at 2 K show no remanent magnetization at zero field. The inset of Fig. \ref{fig:MH_hex} shows the field-dependent magnetization measured at 30 K, in the paramagnetic state, for the two directions. This behavior is linear up to the highest field measured for this temperature, and a linear behavior is also expected for $T>30$ K since the argument of the Brillouin function ($y=g\mu_B H/k_BT$) that describes field-dependent magnetization in the paramagnetic state decreases with increasing temperature. Hence, from this point, the susceptibility for $T\geq 30$ K will be approximated as $\chi\approx M(T,H=10\ \text{kOe})/10\ \text{kOe}$.

The susceptibility for each direction was approximated as $\chi\approx M(T,H=10\ \text{kOe})/10\ \text{kOe}$ and the polycrystalline average was computed as 
\begin{equation}
    \chi_{\text{ave}}=\frac{2\chi_a+\chi_c}{3}.
    \label{eq:polycrystalline}
\end{equation}
The magenta curves in the main panel and inset of Fig. \ref{fig:MH_hex}(a) correspond to the polycrystalline average computed according to Eq. \ref{eq:polycrystalline}. The N\'eel temperature, $T_N=14(2)$~K, is determined as its maximum indicated with a black line in the inset of Fig. \ref{fig:MT_hex}(a). The uncertainty was determined as the temperature range in which $\chi_{ave}$ is less than 5\% of its maximum value, and it is indicated with a gray region in the inset of Fig. \ref{fig:MT_hex}(a). In order to discard any possibility that the subtle nonlinearity of $M(H)$ for a field of 10~kOe applied parallel to $c$ may affect this determination, a more careful estimate of the susceptibility can be done at lower fields, ensuring that it is unambiguously within the linear regime of the $M(H)$ curves plotted in the inset of Fig. \ref{fig:MH_hex}. These results for $H=10$~kOe and $H=100$ Oe are compared in Appendix A, exhibiting minor differences, but ultimately leading to the same estimated value for $T_N$. The value of $T_N$ estimated from magnetization measurements is consistent with that obtained from resistance measurements. 

The inverse susceptibility for the polycrystalline average is plotted in Fig. \ref{fig:MT_hex}(b). A linear fit was performed for the temperature range between 100 and 300~K, in order to obtain the Curie-Weiss parameters indicated in the figure, in accordance with
\begin{equation}
    \chi^{-1}=\frac{\mu_{\text{eff}}^2 N_A}{3k_B}(T-\theta_{CW}),
    \label{eq:CW1}
\end{equation}
where $\mu_{\text{eff}}$ is the effective moment, $N_A$ the Avogadro number, $k_B$ the Boltzmann constant, and $\theta_{CW}$ is the Weiss temperature. An effective moment of $\mu_{\text{eff}}=7.8(1)\ \mu_B$ was obtained for the polycrystalline average, which is consistent with the expected 7.94 $\mu_B$ associated to Eu$^{2+}$.

There is an unexpectedly large anisotropy in the paramagnetic susceptibility, which can be appreciated by comparing the blue and red curves in Fig. \ref{fig:MT_hex}(a). The Eu$^{2+}$ ions, which have a total orbital angular momentum $L=0$, generally exhibit a much smaller single-ion anisotropy given that they do not couple to the crystal electric field. There are other possible sources of anisotropy, such as mixing with excited states with $L\neq 0$ that have been mostly explored in compounds containing Gd \cite{Glazkov2007,Glazkov2005}, or large spin-orbit coupling associated with the six Pt per formula unit, but this is beyond the scope of this work, as we do not have enough information about the magnetic structure in order to address this.

\begin{figure*}
 \centering
 \includegraphics[width=\linewidth]{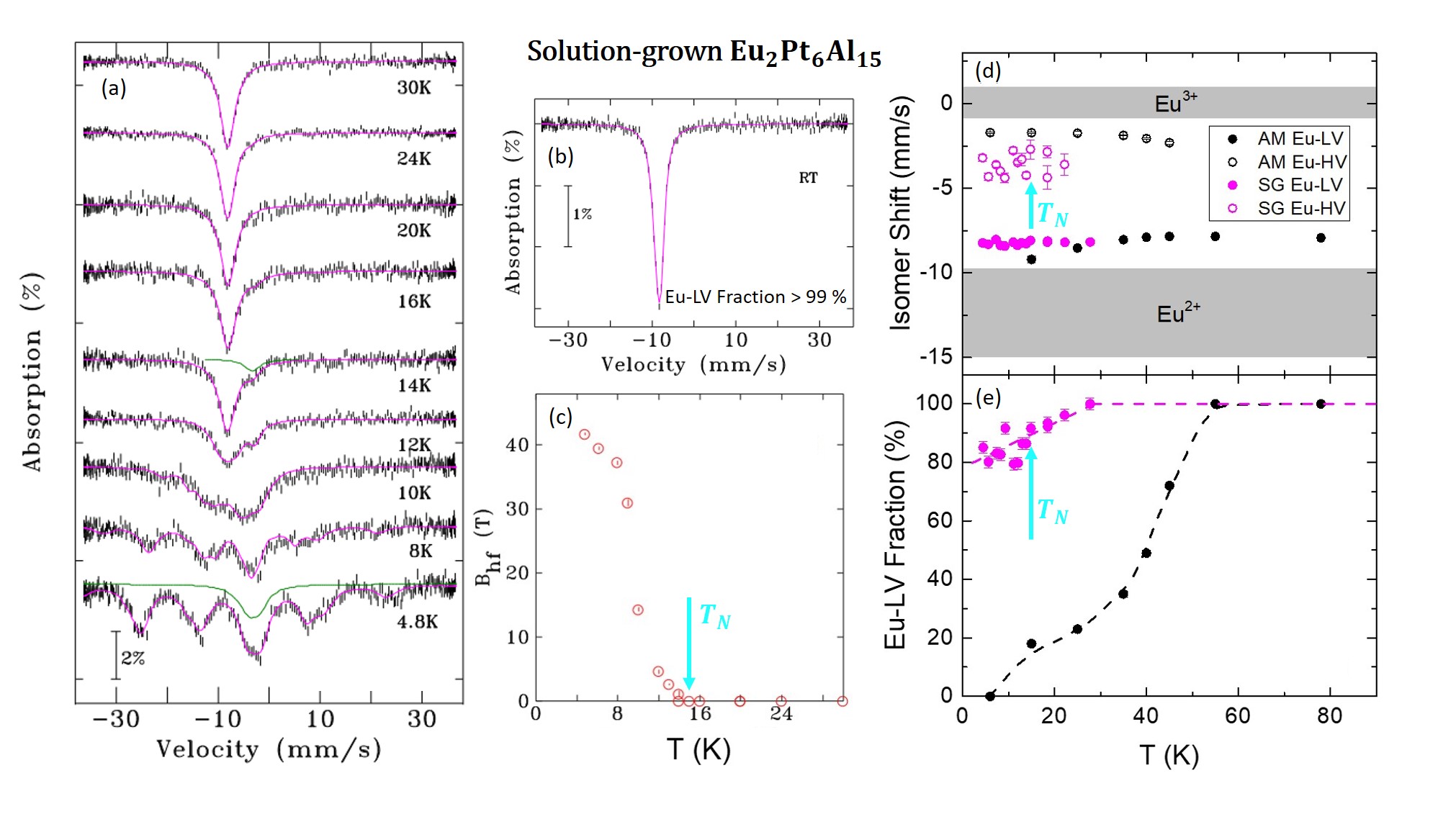}
 \caption{{M\"ossbauer spectroscopy results for the solution-grown (SG) single crystalline Eu$_2$Pt$_6$Al$_{15}$ ground sample: (a) Spectra for different temperatures. The global fit of each spectrum is shown in magenta lines, and the contribution from higher valence Eu (Eu-HV) is shown in green lines for $T=14$~K and 4.8~K, as examples. (b) Room-temperature M\"ossbauer spectrum, showing only low valence Eu (Eu-LV). (c) hyperfine field, $B_{hf}$, as a function of temperature. (d) Isomer shifts of the different Eu valences observed in the Eu$_2$Pt$_6$Al$_{15}$ phase obtained by arc-melting \cite{Radzieowski2018} (AM, black symbols) and in the Eu$_2$Pt$_6$Al$_{15}$ phase obtained from solution growth (SG, magenta symbols). The gray regions indicate the ranges that the isomer shifts take for insulating compounds with Eu$^{2+}$ and Eu$^{3+}$ ions \cite{Steudel2018}. (e) Fraction of the Eu ions with lower valence (Eu-LV) for the arc-melted phase (AM, black symbols) and for the solution grown phase (SG, magenta symbols). The cyan arrows indicate the AFM transition temperature, $T_N$.}}
 \label{fig:Mossbauer}
\end{figure*}

To better understand the valence state of the Eu ions in the solution-grown Eu$_2$Pt$_6$Al$_{15}$, Mössbauer spectra were collected from a ground solution-grown single crystal at different temperatures. Figure \ref{fig:Mossbauer} summarizes these results. The spectra collected at different temperatures and the corresponding global fits are shown in Fig. \ref{fig:Mossbauer}(a) in black and magenta, respectively. The model used for the fits includes two coexisting Eu valences, a lower valence (Eu-LV) and a higher valence (Eu-HV), each with a different isomer shift and a given proportion. The contribution of Eu-HV is explicitly shown in green lines for two selected temperatures ($T=14$ and 4.8~K). The spectrum taken at room temperature, separately shown in Fig. \ref{fig:Mossbauer}(b), is well described by only Eu-LV, and the corresponding fit gives a Eu-HV contibution that is below the 1\% uncertainty. A significant contribution of the latter arises only for $T\leq 24$~K. 

A hyperfine magnetic field, $B_{hf}$, is also added in order to fit the spectra. Figure \ref{fig:Mossbauer}(c) shows the obtained temperature dependence of $B_{hf}$, exhibiting a sharp decrease as temperature increases above 10~K, followed by a slower decrease until reaching zero at 14(1)~K. This is consistent with the AFM order established below 15~K (indicated with the cyan arrows) according to the resistance and magnetic susceptibility measurements. The dependence of $B_{hf}$ does not follow the order-parameter-like behavior typical for second-order antiferromagnetic transitions, likely because the fraction of magnetic Eu-LV is simultaneously changing in the same temperature range. 

The two isomer shifts corresponding to Eu-LV and Eu-HV are plotted in Fig. \ref{fig:Mossbauer}(d) with magenta open and solid symbols, respectively. The ranges corresponding to the isomer shifts of Eu$^{2+}$ and Eu$^{3+}$ for insulating compounds with a clear valence are indicated by the gray regions. The Eu-LV isomer shift of $\sim-8$ mm/s is close to the range corresponding to Eu$^{2+}$ ions but slightly higher. In fact, previous X-ray absorption spectroscopy (XAS) studies, which can probe shorter time scales, indicate that Eu ions in arc-melted Eu$_2$Pt$_6$Al$_{15}$ at room temperature exhibit both Eu$^{2+}$ and Eu$^{3+}$ contributions \cite{Oyama2020}. Similarly, the Eu-HV isomer shift is located around $-4$ mm/s, close to the typical values associated with Eu$^{3+}$ but slightly lower, possibly with some intermediate character as well. The isomer shift of each contribution is independent of temperature, and only their relative intensities change when cooling. Figure \ref{fig:Mossbauer}(e) shows that the fraction of Eu-LV starts decreasing at $\sim$ 24~K, and reaches $\sim$80\% at 4.8~K, leaving $\sim$ 20\% of the Eu in the HV state. It is important to note that the first Eu ions change from the LV state to the HV state at temperatures well above $T_N$, indicated by the cyan vertical arrows in Figs. \ref{fig:Mossbauer}(c), \ref{fig:Mossbauer}(d) and \ref{fig:Mossbauer}(e), but there is only a small reduction in Eu-LV fraction at $T_N$.

For comparison, the parameters reported in Ref. \citenum{Radzieowski2018} for the AM phase are added to both Figs. \ref{fig:Mossbauer}(d) and \ref{fig:Mossbauer}(e) in black open and solid symbols for Eu-LV and Eu-HV, respectively. The main difference between the SG and AM phases is in the fraction of Eu ions in each valence, as shown in Fig. \ref{fig:Mossbauer}(e). For the AM phase, the proportion of Eu-LV starts dropping with the onset of a valence transition around 50~K and eventually approaches 0\% around 5~K, whereas the Eu-LV valence prevails throughout the measured temperature range for the SG phase, dropping only to 80\% at low temperatures. To put this into context, based on the similar high-temperature magnetic susceptibility of the AM and SG samples (compared more in detail in Sec. \ref{subsec:anneal}) we can assume a putative $T_N \gtrsim 15$~K at which the AM would order if all the Eu would remain in the magnetic LV state at low temperatures. However, due to the strong changes in the Eu HV/LV balance in the AM sample, by 15~K there is less than 20\% LV fraction left, an amount that is so depleted of magnetic moment that ordering at $\sim$ 15~K is no longer possible. For 20\% of LV Eu we could expect a magnetic ordering or spin glass behavior at significantly lower temperatures, but by then the LV fraction is reduced to zero (see Fig. \ref{fig:Mossbauer}(e)). For the case of the SG sample, the suppression of the valence transition to low enough temperatures allows the system to establish magnetic ordering. This is reminiscent of PrAl$_2$ which, despite having a putative singlet CEF ground state, the ferromagnetic ordering starts at a temperature that is high enough for Pr to preserve its magnetic moment (due to thermal population of higher CEF states) \cite{Mader1969,OESTERREICHER1977,Palermo1991}.

The case of Eu$M_2X_2$ ($M=$ Rh, Ni, Co; $X=$ Si, Ge) constitutes an even closer example. At ambient pressure, these compounds exhibit nearly divalent Eu ions that order antiferromagnetically, and the application of pressure is able to induce a Eu valence transition above the putative $T_N$ of these compounds, so that the system loses its magnetic moments before they can order \cite{Honda2016,Honda2018}. The case of AM Eu$_2$Pt$_6$(Al$_{1-x}$Ga$_x$)$_{15}$ corresponds to the opposite scenario, as it presents the valence transition for $x=0$, and the negative \textit{chemical pressure} on increasing $x$ was shown to  gradually suppress the valence transition until the Eu ions can adopt an antiferromagnetic (AFM) order at $T_N\sim 15$~K \cite{Oyama2020}. 

In this paper, we show another way to tune the Eu valence transition by varying the synthesis procedure. Only the $c$ lattice parameter shows any significant difference in Tab. \ref{tab:latticeparam} being slightly smaller for the phase with the AFM ordering. The similar suppression of the valence transition and the emergence of magnetic order due to Ga substitution in AM Eu$_2$Pt$_6$(Al$_{1-x}$Ga$_x$)$_{15}$ between $x=0$ and $x=0.1$ is also dominated by comparably slight changes in the $c$ lattice parameter, but in the opposite direction (with a slightly higher $c$ for the antiferromagnetic samples) \cite{Oyama2020}. 
 In the following section, we compare the structures in more detail in order to provide further insight on what causes these changes.

\subsection{Differences in crystal structure}

\begin{figure}[H]
 \centering
 \includegraphics[width=\linewidth]{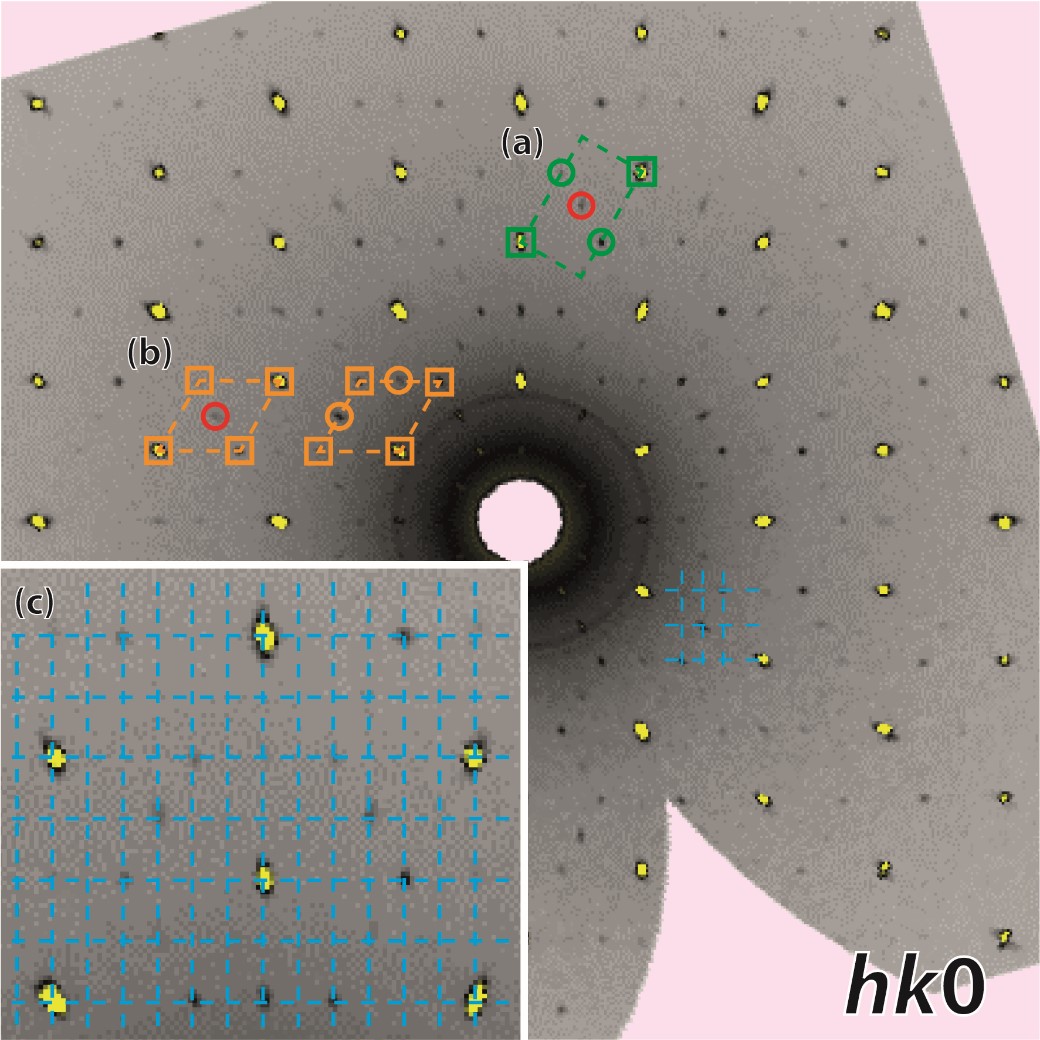}
 \caption{{Section $(hk0)$ of the (reconstructed) diffraction pattern of solution-grown Eu$_2$Pt$_6$Al$_{15}$. The (a) orthorhombic (3+1)D interpretation using one $q_1$ vector (green circles) to describe the satellite reflections is shown in green, and the reflection not matching this description is marked in red. (b) indexation with a hexagonal lattice. The orange dashed line indicates the mesh, the squares mark the main reflections, the circles the satellite reflections, the red circle another mismatched reflection. (c) The blue dashed lines indicate an orthorhombic indexing (see text).}}
 \label{fig:Janka1}
\end{figure}

Since Eu$_2$Pt$_6$Al$_{15}$ has been reported to crystallize in a modulated structure \cite{Radzieowski2018} adopting the Sc$_2$Pt$_6$Al$_{15}$ type structure (approximant in space group $P2_1/m$) \cite{Radzieowski2017}, the reconstructed diffraction images obtained for single crystals of SG Eu$_2$Pt$_6$Al$_{15}$ were investigated in great detail. From the images obtained for the $hk0$ plane shown in Fig. \ref{fig:Janka1}, it becomes readily evident that additional reflections are observed, compared to the initial report of AM Eu$_2$Pt$_6$Al$_{15}$. The originally observed unit cell of AM Eu$_2$Pt$_6$Al$_{15}$ is outlined with the green rectangle (a), the green squares indicate the main reflections justifying the orthorhombic $C$-centered Bravais lattice while the green circles indicate the satellite reflections leading to the interpretation as (3+1)D commensurately modulated structure belonging to the superspace group $Cmcm(\alpha,0,0)0s0$ with $\alpha = 2/3$ and lattice parameters of $a \sim 4.25$, $b \sim 7.30$ and $c \sim 16.30$ \AA. The modulation vector causes an ordering of the Al atoms forming the triangles in the honeycomb Eu substructure with respect to the average structure described in Refs. \citenum{Latturner2002} and \citenum{Niermann2004} before. For further details regarding the analysis and interpretation of the crystal structure of AM Sc$_2$Pt$_6$Al$_{15}$, we refer to the literature \cite{Radzieowski2017}.

It is readily apparent that an additional weak reflection (red circle in green rectangle in Fig. \ref{fig:Janka1}) is observed in the center of the original unit cell. This indicates that, at least from the diffraction point of view, differences between the two structures have to occur. The additional reflection cannot be indexed by the superspace group $Cmcm(\alpha,0,0)0s0$ $(\alpha = 2/3)$ leading to the necessity of a different indexing of the diffraction pattern. One possibility is to index the diffraction pattern with a hexagonal unit cell with $a \sim 7.45$ \AA (orange diamond) and satellite reflections at $q_1 = 0.5 a^*$ and $q_2 = 0.5 b^*$ (orange circles). However, an additional reflection can be observed in some of the hexagonal cells (red circle in orange diamond in Fig. \ref{fig:Janka1}) again contradicting the proposed Bravais lattice. Another possibility is to index the diffraction pattern with the grid pattern shown in blue. In the zoomed region (c), one can clearly see that all observed reflections align with the grid leading to lattice parameters of $a \sim 25.80$ and $b \sim 14.90$ \AA.

\begin{figure}[H]
 \centering
 \includegraphics[width=\linewidth]{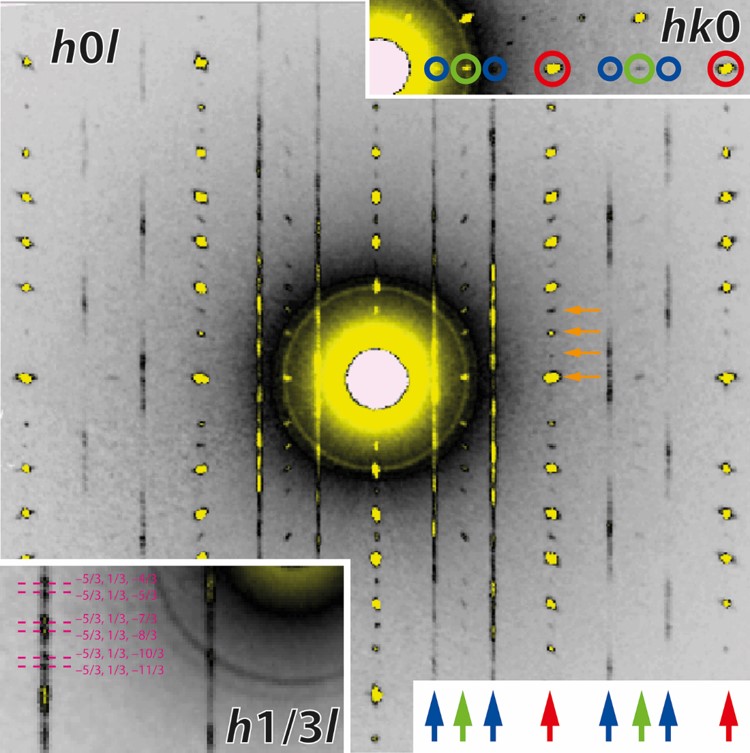}
 \caption{{Section $(h0l)$ of the (reconstructed) diffraction pattern of the solution-grown Eu$_2$Pt$_6$Al$_{15}$. On the top right, a matching section of the $(hk0)$ is shown as inset. The main reflections in line with the unit cell of the arc-melted Eu$_2$Pt$_6$Al$_{15}$ are highlighted via the red and blue circles and red and blue arrows. The additional reflections only visible in solution-grown Eu$_2$Pt$_6$Al$_{15}$ are marked with green arrows and circles. The reflections marked with the blue circles and arrows are somehow diffuse, however, maxima at 1/3 and 2/3 (see zoom at bottom left) are visible.}}
 \label{fig:Janka2}
\end{figure}

To identify whether further differences occur along the crystallographic $c$-axis, the $h0l$ plane (Fig. \ref{fig:Janka2}) must be analyzed. Here, one can clearly identify the intense reflections along $c^*$ (vertical), marked with orange arrows that represent the $c$-axis with $\sim 16.30$ \AA, as observed in AM Eu$_2$Pt$_6$Al$_{15}$. At the top-right of Fig. \ref{fig:Janka2}, a section of the $hk0$ plane is shown that corresponds to the $h0l$ plane shown in large. The circles in red and blue indicate the main reflections as observed in the $hk0$ plane in the AM Eu$_2$Pt$_6$Al$_{15}$, while the green circles indicate the additional reflections in SG Eu$_2$Pt$_6$Al$_{15}$. Even though the $h\ 1/3\ 0$ reflections (blue circles, upper panel) are already present in the AM phase, the SG phase presents additional reflections along the $h\ 1/3\ l$ direction as well (blue arrows). In the zoomed region (lower left), the pink dashed lines indicate maxima on the rather diffuse looking streaks along $c^*$ in an $h\ 1/3\ l$ section, that would lead to a tripling of the $c$-axis in SG Eu$_2$Pt$_6$Al$_{15}$. The unit cell that can be deduced from the reconstructed images is orthorhombic with $a \sim 25.80$, $b \sim 14.90$ and $c \sim 48.90$ \AA$\ $alongside an orthorhombic trilling that is caused by the pseudo-hexagonal symmetry of the diffraction pattern. Indexing the pattern with the software routine led to lattice parameters of $a = 25.838(1)$, $b = 14.911(1)$ and $c = 49.962(3)$ \AA, consistent with a supercell with doubled $a$ and $b$, and tripled $c$. Before proceeding with the interpretation of the structural differences, it has to be pointed out that a potential description of the diffraction pattern with a higher-dimensional superspace group should be possible, however, we were not able to find an appropriate Bravais lattice alongside the matching superspace group. This leads to the problem of the reciprocal space being extremely empty when using the above-mentioned orthorhombic unit cell in comparison to the expected reflections. This in turn causes significant problems during the structure refinement (high $R$-values, significant residual mismatch in electron density). Therefore, we will not present the atomic coordinates or deposit crystallographic data at this stage but rather discuss the differences in the structures of AM and SG Eu$_2$Pt$_6$Al$_{15}$, since differences are clearly visible from the refinement (see below).

\begin{figure}[H]
 \centering
 \includegraphics[width=\linewidth]{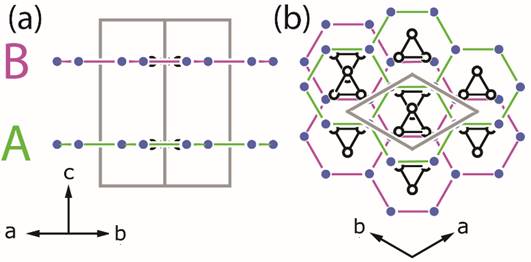}
 \caption{{(a) Honeycomb layers of Eu atoms in the unit cell of arc-melted Eu$_2$Pt$_6$Al$_{15}$ shown perpendicular to the layers. The stacking sequence is given by the capital letters, which are also color coded. (b) Perpendicular view onto the honeycomb layers of Eu atoms along with the Al$_3$ triangles. The [PtAl] slabs in between the layers are omitted for clarity.}}
 \label{fig:Janka3}
\end{figure}

From a structural perspective, the unit cell of solution-grown Eu$_2$Pt$_6$Al$_{15}$ has a tripled $c$-axis, which is indicative of a different stacking along $[001]$. Arc-melted Eu$_2$Pt$_6$Al$_{15}$ (Fig. \ref{fig:Janka3}) consists of slabs of Eu atoms arranged in a honeycomb pattern centered by Al$_3$ triangles (shown in Figure \ref{fig:Janka3}(b)) alternating stacked along the $b$-axis with [PtAl] slabs (not shown). The honeycomb layers are stacked in an …$AB$… sequence. The Al$_3$ triangles within the layers are rotated by $180^{\circ}$ with respect to each other.

\begin{figure}[H]
 \centering
 \includegraphics[width=\linewidth]{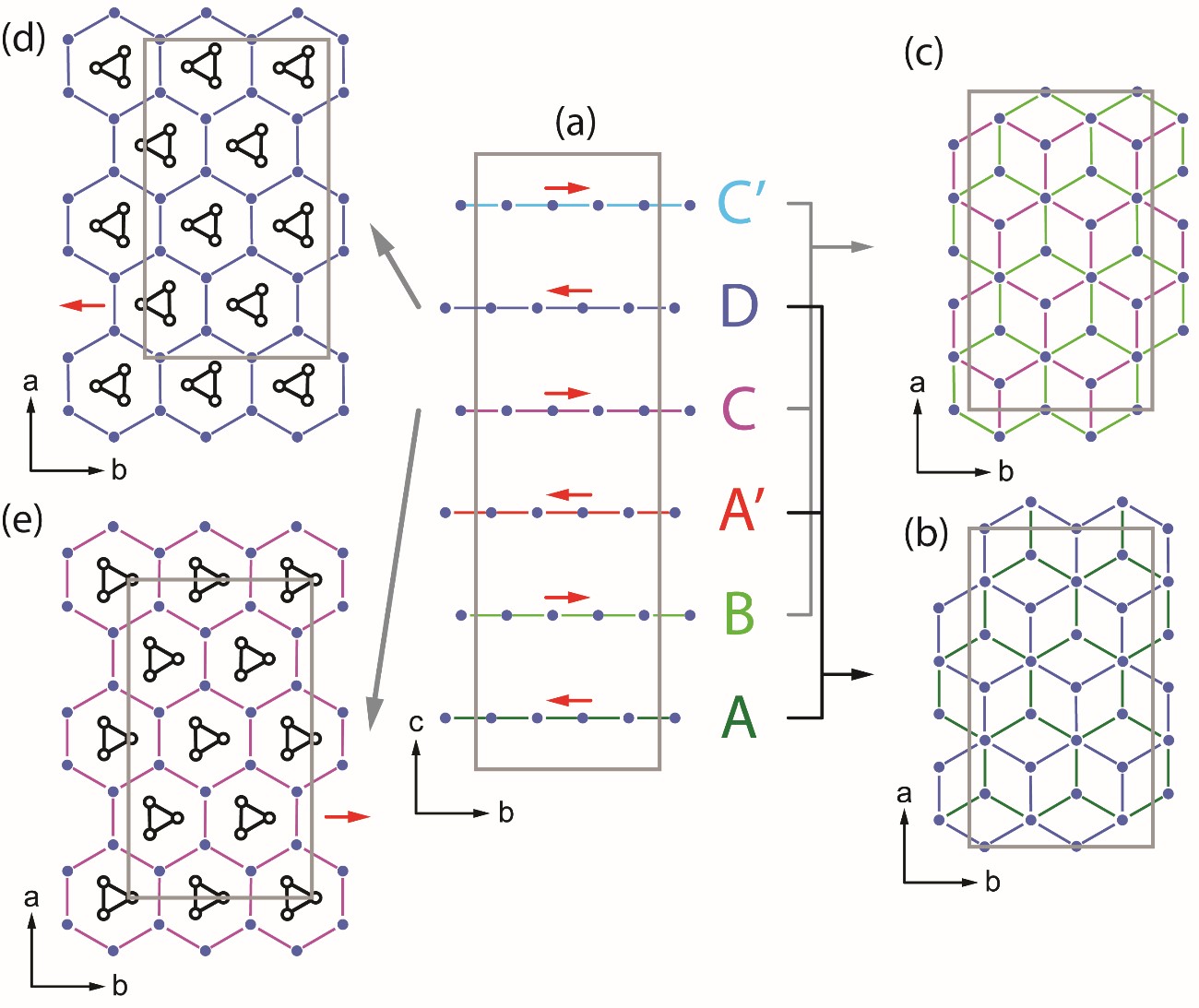}
 \caption{{(a) Honeycomb layers of Eu atoms in the unit cell of solution-grown Eu$_2$Pt$_6$Al$_{15}$ shown perpendicular to the layers along the $a$-axis. The stacking sequence is given by the capital letters, which are also color coded. (b,c) Perpendicular view onto the honeycomb layers of Eu atoms (d,e) along with the Al$_3$ triangles. The [PtAl] slabs in between the layers are omitted for clarity.}}
 \label{fig:Janka4}
\end{figure}

In solution-grown Eu$_2$Pt$_6$Al$_{15}$ the stacking sequence is three times larger, reflected in the $3c$ unit cell. Note that the stacking sequence is along the $b$-axis in the arc-melted, but along the $c$-axis in the flux-grown sample (different unit cell setting). Now, a stacking sequence of …$ABA’CDC’$… is observed as shown in Figure \ref{fig:Janka4}(a). The color coding of the respective layers visualizes that $A$ and $A’$ have the same orientation, so the red layer $A’$ is not visible in the representation along $[001]$ shown in Figure \ref{fig:Janka4}(b). The same is true for $C$ and $C’$ since the light blue layer $C’$ is directly on top of the dark blue layer $C$ in Fig. \ref{fig:Janka4}(c). Figures \ref{fig:Janka4}(d) and \ref{fig:Janka4}(e) finally visualize the arrangement of the Al$_3$ triangles which are oriented either along $-b$ (minus $b$) (Fig. \ref{fig:Janka4}(d)) or along $b$ (Fig. \ref{fig:Janka4}(e)). The orientation is labeled by the red arrows in Figs. \ref{fig:Janka4}(a), \ref{fig:Janka4}(d) and \ref{fig:Janka4}(e).

The representation shown in Figure \ref{fig:Janka4}, however, does not explain the doubling of the unit cell along $a$ and $b$ with respect to arc-melted Eu$_2$Pt$_6$Al$_{15}$. Therefore, this enlargement needs to be caused by the [PtAl] slabs. The refinements only show that there are slight differences in the Pt–Al distances, but no clear structural distortion is visible. This further strengthens the idea that a description as a modulated structure via the superspace approach should be more feasible here.

\subsection{Effects of thermal treatments}
\label{subsec:anneal}

Given the clear physical and structural differences between the SG single crystals and the AM samples, we performed several annealing and synthesis studies.

\begin{figure}[H]
 \centering
 \includegraphics[width=\linewidth]{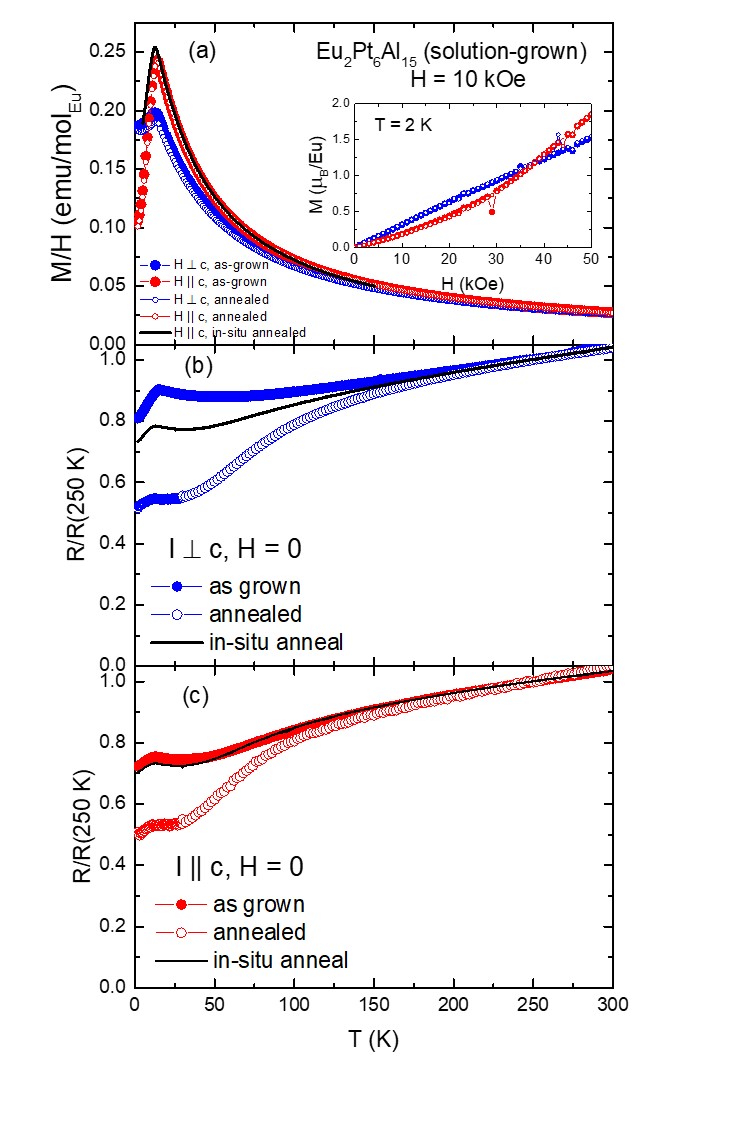}
 \caption{{(a) Temperature-dependent (main panel) and field-dependent (inset) magnetization of the solution-grown Eu$_2$Pt$_6$Al$_{15}$ with the field applied perpendicular (blue) and parallel (red) to $c$, for the as-grown crystals (in solid symbols), for the crystals annealed at 600 $^{\circ}$C (open symbols), and for the crystals annealed in-situ at 900 $^{\circ}$C (black line). (b) Normalized resistance perpendicular to $c$ for the as-grown (solid symbols), annealed (open symbols), and in-situ annealed (black line). (c) Normalized resistance along $c$ for the as-grown (solid symbols), annealed (open symbols), and in-situ annealed (black line).}}
 \label{fig:anneal}
\end{figure}

The first thermal treatment attempted on the solution-grown Eu$_2$Pt$_6$Al$_{15}$ single-crystals consisted on a 600 $^{\circ}$C anneal for 86 hours. This was done by selecting a few crystals and sealing them in a silica tube with a partial pressure of argon and placing them in a box furnace for annealing, after which the furnace was turned off and the ampoule was allowed to slowly cool inside the furnace over roughly 12 hours. As shown with open symbols in Fig. \ref{fig:anneal}(a), this thermal treatment did not have any major effect on the magnetic properties of the system. However, it did result in a modest increase of the RRR for both directions, as shown in Figs. \ref{fig:anneal}(b) and \ref{fig:anneal}(c), consistent with a reduction of the defect scattering due to the anneal. The second thermal treatment consisted of an anneal for the same time at 750 $^{\circ}$C which lead to the decomposition of the sample surface, possibly due to the melting of flux droplets on the surface of the crystal.

In order to prevent this degradation of the crystals when annealing at temperatures higher than the melting temperature of the flux, we attempted another type of thermal treatment which we refer to as \textit{in situ anneal}. This consisted of repeating the procedure to grow the single-crystals, but adding an additional step at the end: dwelling at the final temperature (900 $^{\circ}$C) for 100 hours before decanting the excess solution. This way, the crystals are already in equilibrium with the remaining liquid at high temperatures, reducing the risk of decomposition of the crystal surface. The results measured for the in-situ annealed samples are shown in black lines in Figs. \ref{fig:anneal}(a), \ref{fig:anneal}(b) and \ref{fig:anneal}(c), displaying even less changes than the anneal done at 600 $^{\circ}$C. This indicates that temperatures lower than 900 $^{\circ}$C are more effective in curing the defects in the crystal, as reflected by the smaller change in the RRR of the in situ annealed samples. The important point made by all of these thermal studies is that, whereas we can improve RRR slightly, we cannot change the magnetization in any significant manner. This suggests that the new structure we found for the SG sample is the result of slowly growing the crystal over temperature ranges higher than 900 $^{\circ}$C, rather than quenching from a higher-temperature melt, as is the case of the AM sample. 

\begin{figure}[H]
\centering
\includegraphics[width=\linewidth]{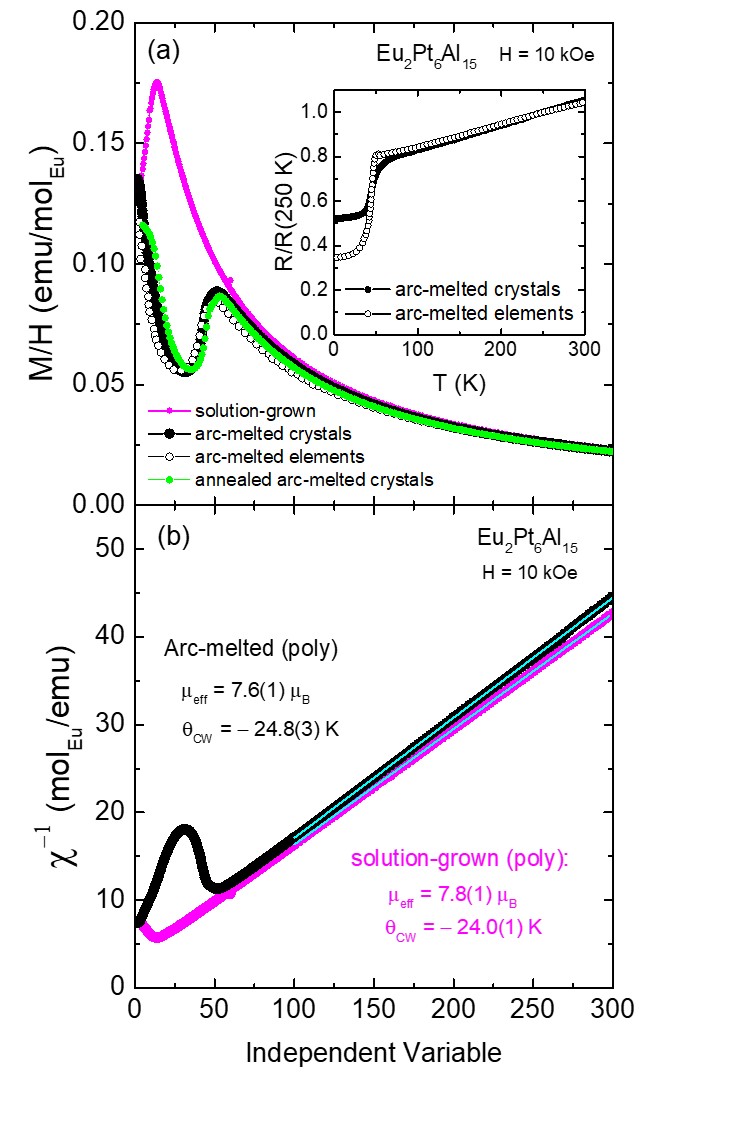}
\caption{{(a) Main panel: polycrystalline average magnetization of the solution-grown Eu$_2$Pt$_6$Al$_{15}$ (solid magenta symbols), magnetization of the arc-melted crystals before (solid black symbols) and after annealing (solid green symbols), and magnetization of the reported \cite{Radzieowski2018} arc-melted polycrystalline sample (open black symbols) normalized by the applied field of 10~kOe, as a function of temperature. Inset: temperature-dependent resistance of a piece of the arc-melted crystals (solid black symbols) and the reported for the arc-melted phase (open black symbols), normalized by their values at 250~K. (b) Main panel: inverse susceptibility as a function of temperature for the polycrystalline average of the solution-grown Eu$_2$Pt$_6$Al$_{15}$ (magenta) and for the arc-melted crystal (black); the linear fits are indicated with cyan dashed lines.}}
\label{fig:arcmelt_hex}
\end{figure}

Given that none of the thermal treatments we tried led to the stabilization of the reported Eu$_2$Pt$_6$Al$_{15}$ phase obtained by arc-melting the constituting elements together \cite{Radzieowski2018}, we took SG crystals and arc-melted them. After the crystals were arc-melted, the temperature-dependent magnetization and resistance were measured on pieces of the obtained polycrystalline button, shown with black solid symbols in the main panel and inset of Fig. \ref{fig:arcmelt_hex}(a), respectively. These reproduce the behavior originally reported in Ref. [\citenum{Radzieowski2018}], shown with black open symbols. Larger discrepancies are observed in the resistance, probably due to a larger concentration of defects in the sample measured in this paper compared to the reported one, resulting in a smaller RRR. The magnetization results display a better agreement with the reported behavior, which is drastically different from the behavior of the solution-grown single crystals shown with magenta symbols (for the polycrystalline average) in the main panel of Fig. \ref{fig:arcmelt_hex}(a) for comparison. 

A 600 $^{\circ}$C anneal was also performed for the AM Eu$_2$Pt$_6$Al$_{15}$ for 86 hours, following the same procedure as described for the SG Eu$_2$Pt$_6$Al$_{15}$ crystals. No significant changes where induced in the magnetic properties of the sample, as shown in green symbols in Fig. \ref{fig:arcmelt_hex}(a).

The inverse susceptibility shown in black in Fig. \ref{fig:arcmelt_hex}(b) was also fitted for the range $100\ \text{K}\leq T\leq 300$~K, yielding an effective moment of $7.6(1)\ \mu_B$ and a Weiss temperature of $-24.8(3)$~K, both of which are not far from the polycrystalline values of the solution-grown phase. The similarity of the Weiss temperatures obtained from the high-temperature SG and AM phases suggest that the AM phase would also order antiferromagnetically at a similar putative $T_N$ as the SG phase if the Eu would not lose its magnetic moment due to the valence transition occurring at higher temperatures.
 
\begin{figure}[H]
\centering
\includegraphics[width=\linewidth]{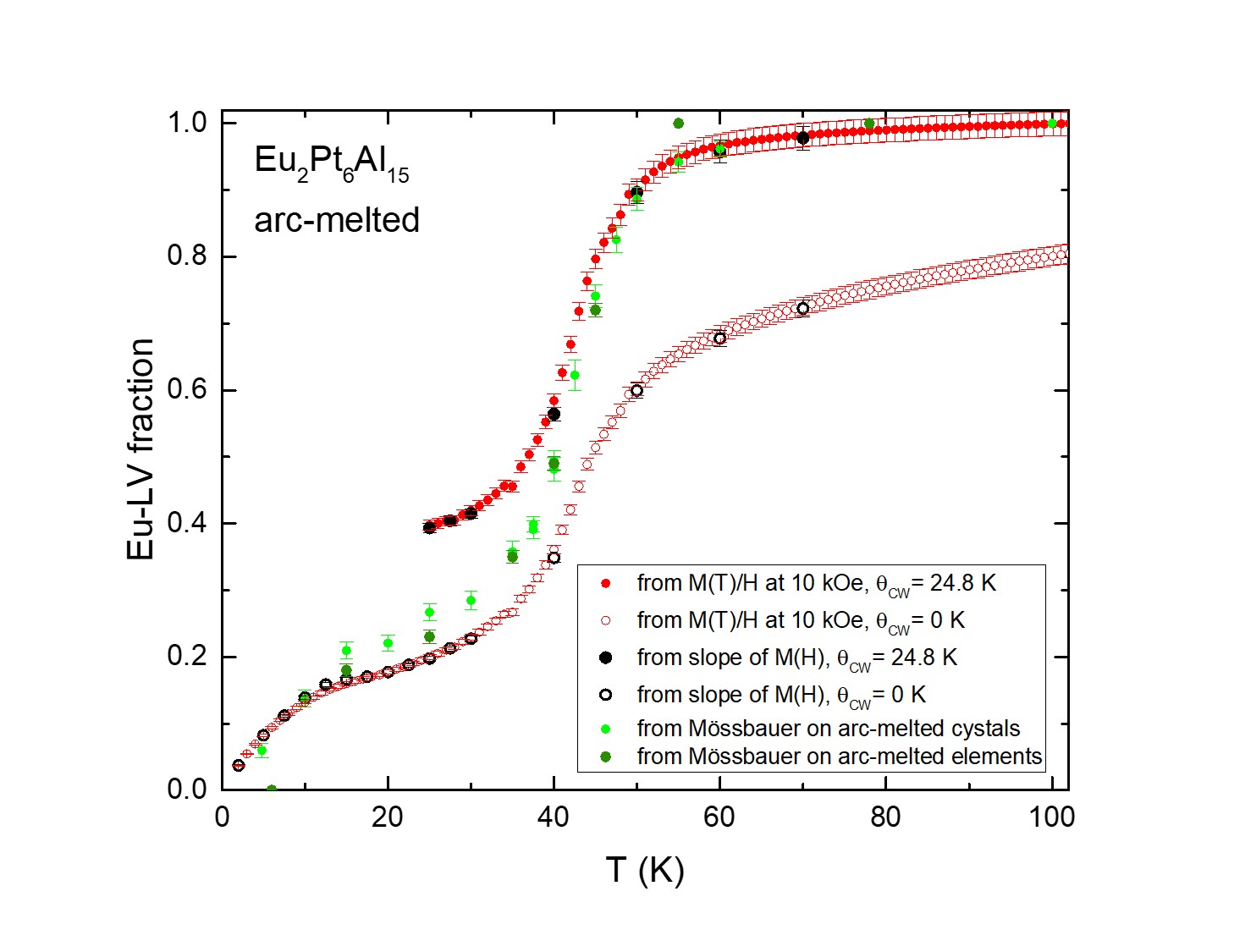}
\caption{{Temperature dependence of the Eu-LV fraction determined from Mössbauer measurements on the arc-melted crystals (solid light-green symbols) as well as on the reported \cite{Radzieowski2018} phase obtained by arc-melting the elements (solid dark-green symbols). The plot also includes the $M(T)$ measurements assuming the high-temperature (solid red symbols) and low-temperature (open red symbols) limits, and the slope in $M(H)$ measurements assuming the high-temperature (solid red symbols) and low-temperature (open red symbols) limits.}}
\label{fig:Mossbauer_MT_MH}
\end{figure}

The substantial low-temperature upturn in the magnetization of the arc-melted phase shown in Fig. \ref{fig:arcmelt_hex}(a) could be attributed either to the presence of paramagnetic impurities or to the fact that there is no abrupt change in the valence of all the Eu ions as soon as the temperature crosses 50~K, but rather a coexistence of Eu-LV and Eu-HV ions below this temperature. In order to evaluate this in greater detail, we also measured Mössbauer spectra for the phase obtained after arc-melting the solution-grown crystals, in order to estimate the fraction of Eu-LV as a function of temperature. The results of our measurements are plotted with green open symbols in Fig. \ref{fig:Mossbauer_MT_MH}. 

The fraction, $x$, of Eu-LV can be related to the magnetic susceptibility through the following expression
\begin{equation}
    \chi=x\frac{C_{LV}}{T-\theta_{LV}}+(1-x)\frac{C_{HV}}{T-\theta_{HV}},
\end{equation}
where $C_{LV}$ and $C_{HV}$ correspond to the Curie constants associated with the Eu-LV and Eu-HV, respectively; while $\theta_{LV}$ and $\theta_{HV}$ correspond to their Curie temperatures. If we neglect the contribution of the lower-moment Eu-HV to the susceptibility, since it is closer to the nonmagnetic Eu$^{3+}$, we are left with
\begin{equation}
    \chi(T)\approx x(T)\frac{C_{LV}}{T-\theta_{LV}(x(T))}.
    \label{eq:CW_3}
\end{equation}
The value of $\theta_{LV}$ is directly related to the exchange coupling between the Eu-LV ions, which has a non-trivial dependence with temperature as the fraction of Eu-LV changes. As the fraction of these magnetic ions changes, so does the distance between them. The exchange coupling, governed by the RKKY interaction, is sensitive to the changing distance between the magnetic ions as $x$ changes. However, we can assume that, in the low-temperature limit where $x\rightarrow 0$, the Eu-LV ions are so far from each other that their coupling strength will also approach zero. On the other hand, for high enough temperatures such that $x\rightarrow 1$, the spatial distribution of the Eu-LV ions will be similar to that in the paramagnetic state, for which we can use the Curie temperature estimated from the Curie-Weiss fit in Fig. \ref{fig:arcmelt_hex}(b). In summary, we can approximate Eq. \ref{eq:CW_3} in those limits as
\begin{equation}
    \chi(T)\approx\begin{cases} 
x(T)\frac{C_{LV}}{T} & \text{when } T\rightarrow0 \\
x(T)\frac{C_{LV}}{T+24.8\ \text{K}} & \text{when } T\rightarrow 45\ \text{K}.
\end{cases}
\end{equation}
The temperature-dependent susceptibility was estimated, on the one hand, by dividing the temperature-dependent magnetization $M(T)$ by the applied field $H$. The solid and open red symbols in Fig. \ref{fig:Mossbauer_MT_MH} correspond to the estimated Eu-LV fraction, $x$, using that susceptibility in the high and low-temperature limit, respectively. On the other hand, $\chi$ can be obtained more accurately by performing $M(H)$ measurements at different temperatures, and evaluating the slopes at fields between $1\ \text{kOe}\leq H\leq 5\ \text{kOe}$. The black solid and open symbols in Fig. \ref{fig:Mossbauer_MT_MH} correspond to the estimated Eu-LV fraction, $x$, using that susceptibility in the high and low-temperature limit, respectively.

The Eu-LV fraction obtained by Mössbauer measurements approaches the corresponding limits at high and low temperatures. This is an indication that the low-temperature upturn of the magnetic susceptibility is indeed explained by the coexistence of magnetic Eu-LV and (almost) non-moment-bearing Eu-HV ions, without the need to invoke paramagnetic impurities in the system. Since the magnetic susceptibility measured on a piece of the arc-melted button is consistent with the Mössbauer results measured on powder, we can conclude that the process of grinding is not significantly altering the magnetic behavior of the samples. Further more, Fig. \ref{fig:Mossbauer_MT_MH} shows that the Eu-LV fraction obtained for the arc-melted crystals (light-green symbols) reproduces well the reported one \cite{Radzieowski2018} for the phase obtained by arc-melting the elements together (dark-green symbols).

Based on our conjecture that the quenching from a higher-temperature during the synthesis could lead to the discussed differences in structure and physical properties, we modified the solution growth procedure for the crystals to grow at higher temperatures. The obtained single crystals had the same chemical composition and similar crystal structure but exhibited intermediate properties to the SG and AM phases described above. As a result, we find further evidence that we can tune the valence transition temperature relative to the antiferromagnetic ordering temperature. Details on the growth and characterization of this phase are provided in Appendix B.

\section{Conclusion}
\label{sec:Conclusion}

Despite the fact that the discovered Eu$_2$Pt$_6$Al$_{15}$ phase has the same composition as a previously reported phase obtained by arc-melting \cite{Radzieowski2018}, the single crystal X-ray diffraction measurements revealed that the former is a superstructure of the latter, with doubled $a$ and $b$ lattice parameters, and tripled $c$ lattice parameter. Both structures contain the same slabs of honeycomb layers of Eu centered by Al$_3$ triangles, alternated with [PtAl] slabs. However, in the solution-grown phase, the slabs are stacked in an $...ABA'CDC'...$ sequence, instead of the $...AB...$ sequence reported for the phase obtained by arc-melting. 

This subtle difference in structure results in strong differences in the valence state of the Eu ions as well as the sample's low temperature electronic and magnetic properties. On the one hand, the parent phase obtained by arc-melting presents a valence transition at around 50~K in which the Eu ions lose their magnetic moment. On the other hand, in the polymorph obtained by solution growth, all the Eu ions preserve their magnetic moments down to 30~K, below which only a small fraction of them change their valence. In essence, the change in stacking of the layers in the superstructure decreases the valence transition to low enough temperatures, eventually allowing the Eu moments to order antiferromagnetically below 15~K and establish an internal field that rapidly increases to a value of $\sim40$ T, as shown in Fig. \ref{fig:Mossbauer}(c). The observation of an antiferromagnetic transition relies on the relative values of the putative ordering temperature, $T_N$, and the valence transition temperature, $T_V$. The situation is to some extent analogous to the case of PrAl$_2$, in which the ordering temperature needs to be higher than the crossover to the Pr singlet ground state in order to exhibit ferromagnetism \cite{Mader1969,OESTERREICHER1977,Palermo1991}.

Previous cases have demonstrated the tunability of the Eu valence transition above or below a putative antiferromagnetic transition: by applying hydrostatic pressure on Eu$M_2X_2$ ($M=$ Rh, Ni, Co; $X=$ Si, Ge) \cite{Honda2016,Honda2018}, or by changing the chemical composition in the arc-melted Eu$_2$Pt$_6$(Al$_{1-x}$Ga$_x$)$_{15}$ \cite{Oyama2020}. In this work we have shown that minor changes in the crystal structure, such as the way the layers are stacked, can also strongly tune this transition. We are able to control this by modifying the synthesis procedures in which Eu$_2$Pt$_6$Al$_{15}$ is obtained. We conjecture that growing the crystals by cooling down slowly to lower temperatures can favor the more complex $...ABA'CDC'...$ stacking which suppresses the valence transition, as opposed to quenching the crystals from much higher temperatures after arc-melting which favors a simpler $...AB...$ stacking. We were able to obtain crystals of some intermediate phase by adapting the initial composition and the temperature profiles so that the crystals grow at higher temperatures. Further $^{151}$Eu M\"ossbauer and single crystal X-ray studies are needed to understand the Eu valence behavior and the detailed crystal structure of this variant. 

The Eu$_2$Pt$_6$Al$_{15}$ phase obtained by solution-growth preserved its properties after different annealing treatments. However, by arc-melting the solution-grown crystals, we obtained the same phase that had previously been synthesized by arc-melting the elemental solids together \cite{Radzieowski2018}. $^{151}$Eu M\"ossbauer measurements done on this phase were consistent with the temperature dependent magnetization results, indicating that the low-temperature upturn in the susceptibility is most probably not due to paramagnetic impurities.\\

\section{Acknowledgements}
The authors acknowledge Rebecca Flint and Shuyuan Huyan for useful discussions. Work at Ames National Laboratory (J.S., S.H., C.L.M, A.S., R.F.S.P., T.J.S., S.L.B., and P.C.C.) was supported by the U.S. Department of Energy, Office of Basic Energy Science, Division of Materials Sciences and Engineering. Ames National Laboratory is operated for the U.S. Department of Energy by Iowa State University under Contract No. DE-AC02-07CH11358. Work at McGill University (D.H.R.) was supported by the Fonds Québécois de la Recherche sur la Nature et les Technologies and the Natural Sciences and Engineering Research Council (NSERC) Canada. DHR benefits from his affiliation to the Regroupement Québécois sur les Matériaux de Pointe (RQMP) https://doi.org/10.69777/309032. Work at Universitat des Saarlandes (O.J.) was supported by the German Research Foundation DFG (JA 1891-10-1). R.F.S.P. was also supported by the São Paulo Research Foundation (FAPESP), Brazil, process Number 2024/08497-6.

\section{Data availability}

The data that support the findings of this article are posted and available \cite{data_Eu2Pt6Al15}.

\section{Appendix A: low-field susceptibility}
\label{AppendixA}

\begin{figure}[H]
 \centering
 \includegraphics[width=\linewidth]{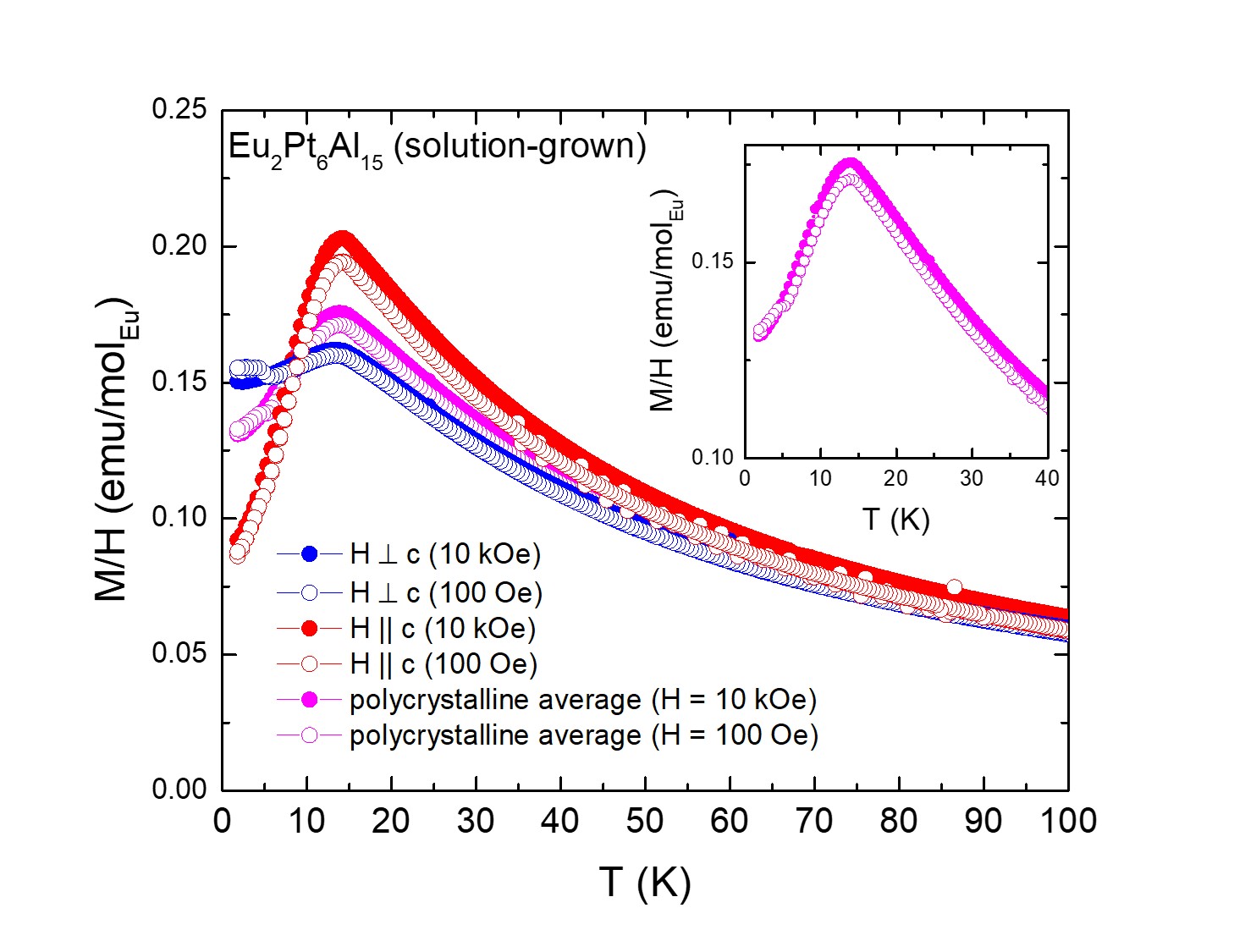}
 \caption{{Main panel: Temperature dependent ZFC and FC magnetization by the applied field of 100 Oe of the solution-grown Eu$_2$Pt$_6$Al$_{15}$ phase with the field applied perpendicular (blue) and parallel (red) to $c$, as well as for the polycrystalline average (magenta). Inset: the $d(\chi_{\text{ave}}T)/dT$ as a function of temperature for a lower temperature range. The green and black lines indicate the criteria by which the transition temperature and its error were estimated.}}
 \label{fig:MT_hex_100Oe}
\end{figure}

Figure \ref{fig:MT_hex_100Oe} shows the temperature-dependent magnetization normalized by the applied field of 100 Oe applied perpendicular (blue open symbols) and parallel (red open symbols) to the $c$ axis, compared to those measured at 10~kOe (solid symbols), for SG Eu$_2$Pt$_6$Al$_{15}$. They were measured under ZFC and FC protocols which perfectly overlap with each other, and are consistent with antiferromagnetic ordering, with the magnetic moments oriented parallel to $c$. The polycrystalline average of the magnetic susceptibility is represented with magenta symbols in the main panel of Fig. \ref{fig:MT_hex_100Oe}. The $d(\chi_{\text{ave}} T)/dT$ are plotted in the inset. The intersections between this line and the black lines that extrapolate the low- and high-temperature behaviors were taken, and the middle point was chosen to be the $T_N$, while the error bar was defined as half of the distance between the two intersections. The  feature appearing at $T\sim 5$~K probably corresponds to an impurity and not to the studied phase, since the feature was drastically increased after the anneal done at 750 $^{\circ}$C that degraded the sample surface, and later suppressed when polishing the sample.

\section{Appendix B: Single-crystal growth at higher temperatures}
\label{AppendixB}

Based on our conjecture that the quenching from a higher-temperature during the synthesis could lead to the discussed differences in structure and physical properties, we modified the solution growth procedure for the crystals to grow at higher temperatures (SG-HT). We chose the initial composition of the melt to be Eu$_6$Pt$_{19}$Al$_{75}$, that is closer to the stoichiometry of the target Eu$_2$Pt$_6$Al$_{15}$ phase, with a reduced excess of Al. This was heated to 1200 $^{\circ}$C and kept at that temperature for 10 hours, followed by a slow cooling to 1190 $^{\circ}$C over the course of 48 hours. Once this final temperature was reached, the excess solution was decanted.

\begin{figure}[H]
 \centering
 \includegraphics[width=\linewidth]{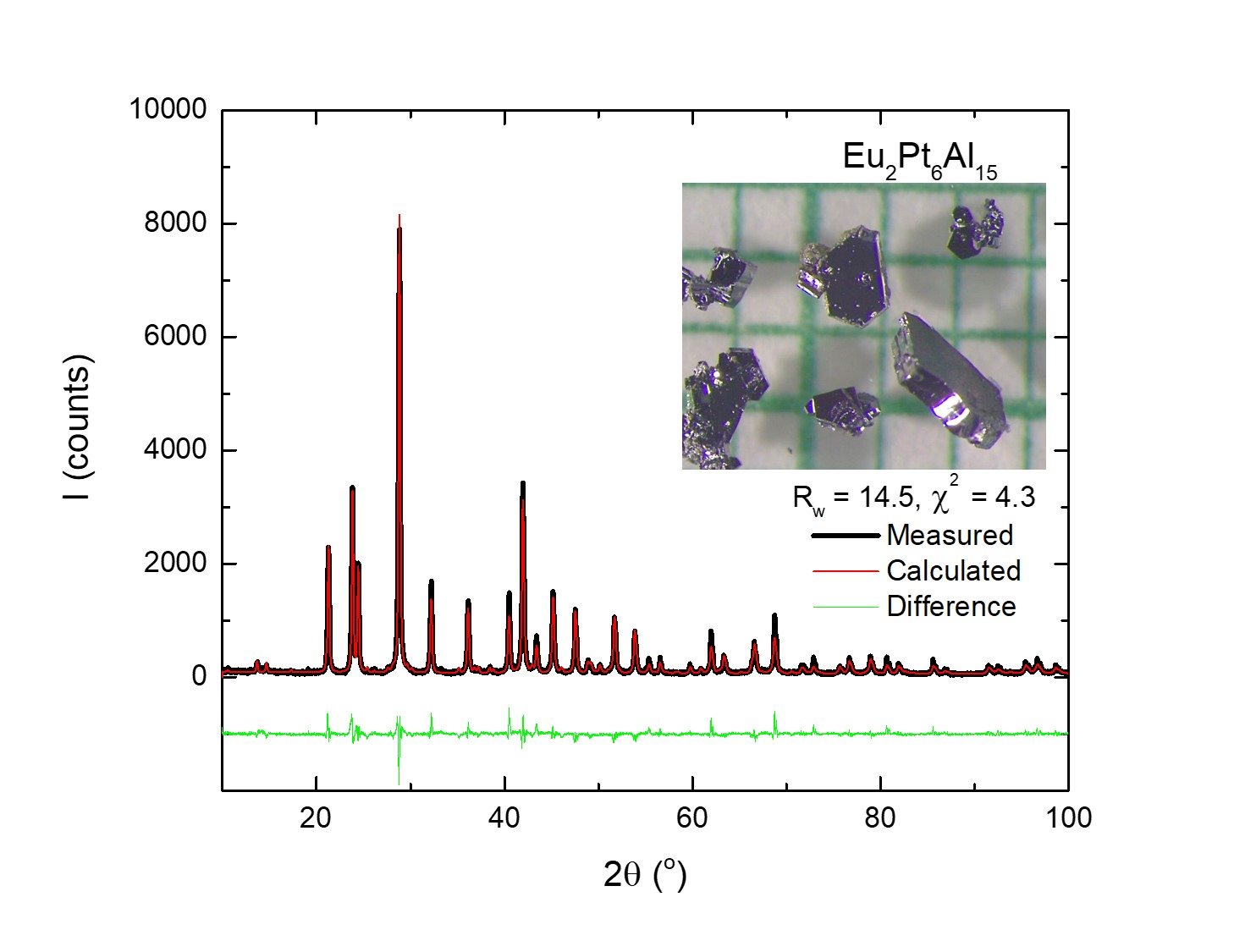}
 \caption{{(a) Powder X-ray diffraction pattern of the Eu$_{2}$Pt$_{6}$Al$_{15}$ phase obtained by the modified growth method (black), the best fit obtained by Rietveld refinement (red) and the residues (green). A photograph of two crystals is also shown.}}
 \label{fig:pxrd_intermediate}
\end{figure}

Powder XRD and EDS confirmed that the Eu$_2$Pt$_6$Al$_{15}$ was obtained. The measured fractions of Eu, Pt and Al were 0.085(2), 0.27(1) and 0.64(8), respectively. The results of the Rietveld refinement are shown in Fig. \ref{fig:pxrd_intermediate}, which yielded the lattice parameters $a=7.45(2)$ \AA, $b=7.44(2)$ \AA, $c=16.650(1)$ \AA, and $\beta=119.96(3)^{\circ}$.

\begin{figure}[H]
 \centering
 \includegraphics[width=\linewidth]{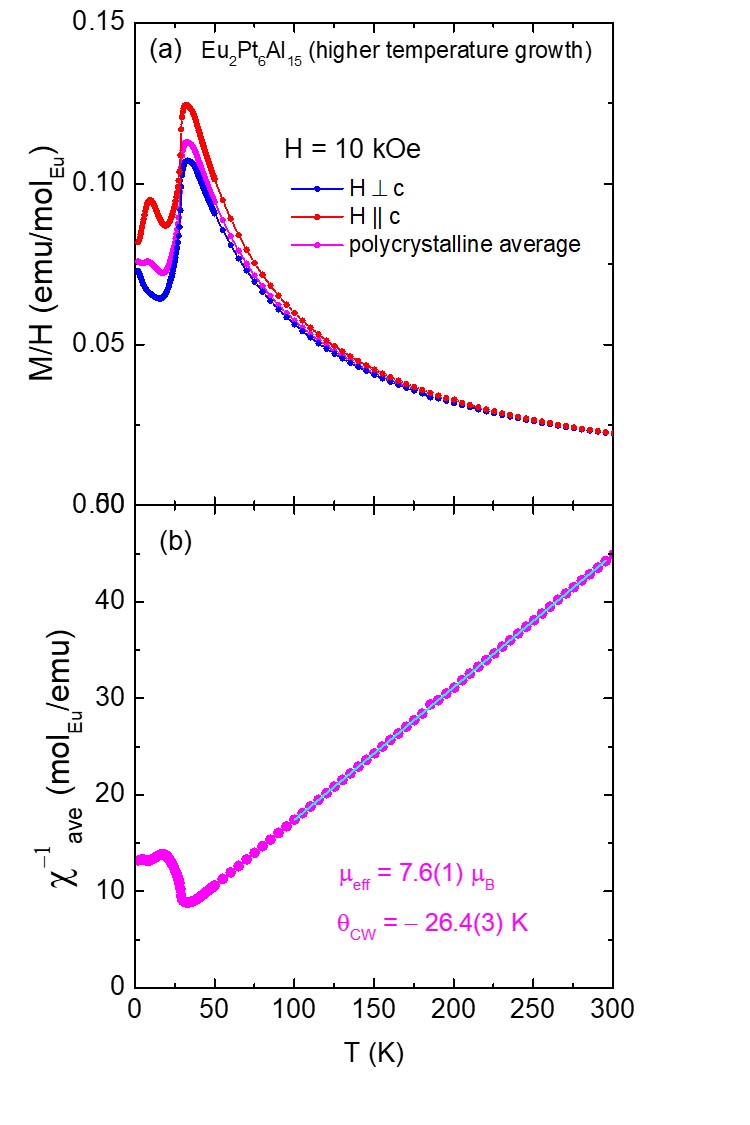}
 \caption{{(a) Temperature-dependent ZFC and FC magnetization divided by the applied field of 10 kOe of the Eu$_2$Pt$_6$Al$_{15}$ phase grown above 1190 $^{\circ}$C, with the field applied perpendicular (blue) and parallel (red) to $c$, as well as for the polycrystalline average (magenta). (b) Polycrystalline average of the inverse susceptibility as a function of temperature for the same phase (magenta); the linear fit is indicated with a cyan line.}}
 \label{fig:CW_intermediate}
\end{figure}

Figure \ref{fig:CW_intermediate}(a) shows the temperature-dependent magnetization  measured under a field of 10 kOe applied perpendicular (blue) and parallel (red) to $c$, as well as the corresponding polycrystalline average (magenta).  The polycrystalline-averaged inverse susceptibility is plotted in Fig. \ref{fig:CW_intermediate}(b), on which a Curie-Weiss fit was performed. This phase has an effective moment of 7.6(1) $\mu_B$, which is consistent with both the arc-melted crystals (AM-Crys) and the solution-grown crystals decanted at lower temperatures (SG-LT), but is closer to the former (7.64(1) $\mu_B$).

\begin{figure}[H]
 \centering
 \includegraphics[width=\linewidth]{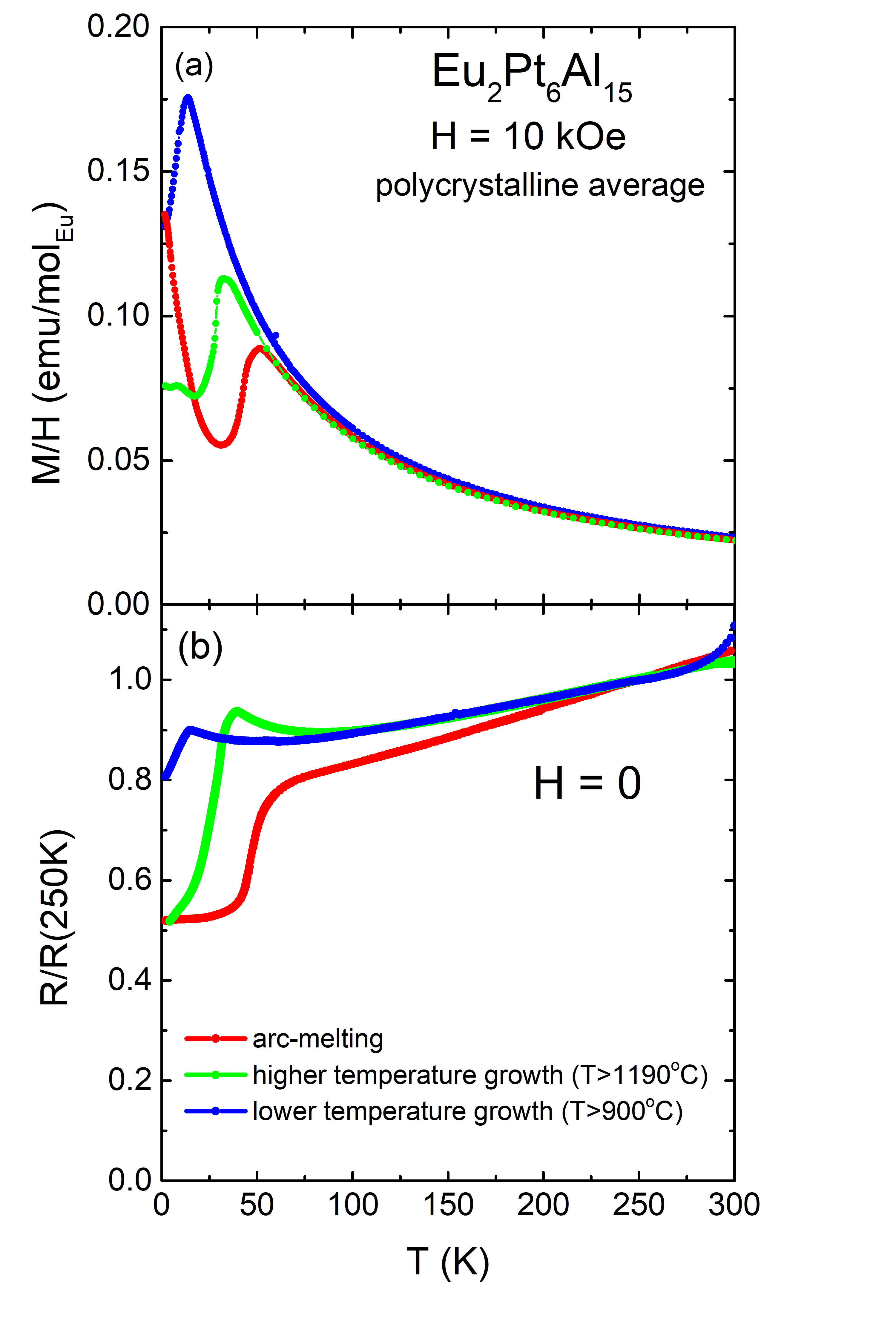}
 \caption{{(a) Temperature-dependent ZFC and FC polycrystalline-averaged magnetization divided by the applied field of 10 kOe of the Eu$_2$Pt$_6$Al$_{15}$ phase obtained by arc-melting (red), by solution growth above 1190 $^{\circ}$C (green) and by solution growth above 900 $^{\circ}$C. (b)  Temperature-dependent resistance normalized by its value at 250 K, measured upon cooling and warming (no observable difference) of the Eu$_2$Pt$_6$Al$_{15}$ phase obtained by arc-melting (red), by solution growth above 1190 $^{\circ}$C (green) and by solution growth above 900 $^{\circ}$C.}}
 \label{fig:comparison}
\end{figure}

Figure \ref{fig:comparison} shows a comparison of the polycrystalline-averaged susceptibilities (panel a) and the normalized resistances (panel b) of the arc-melted crystals (AM-Crys) with the two phases grown in single-crystalline form (SG-HT and SG-LT). The SG-HT (green curves) displays a drop in the magnetization and resistance occurring at a lower temperature than the valence transition in the AM-Crys sample, but at a higher temperature than the magnetic transition in the SG-LT sample. 

\begin{figure}[H]
 \centering
 \includegraphics[width=0.9\linewidth]{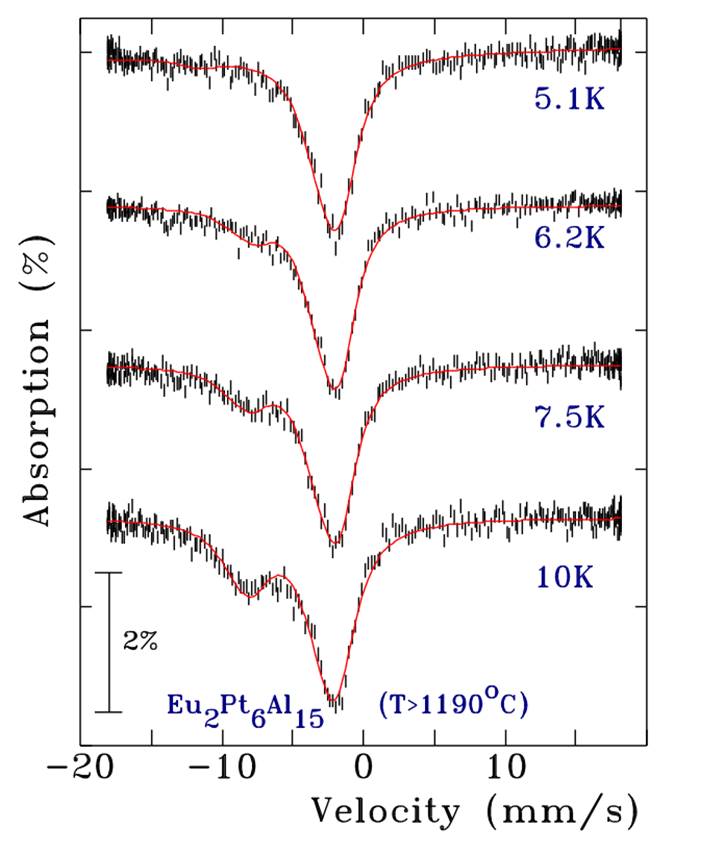}
 \caption{{Spectra for different temperatures ($5.1\leq T\leq 10$~K) for the high-temperature solution-grown Eu$_2$Pt$_6$Al$_{15}$ sample. The global fit of each spectrum is shown in red lines.}}
 \label{fig:Mossbauer_spectra}
\end{figure}

M\"ossbauer measurements were done at temperatures $ 5.1\leq T\leq 50 $ K, in order to clarify the nature of the transition observed in the SG-HT sample. Figure \ref{fig:Mossbauer_spectra} shows the M\"ossbauer spectra measured for the lowest temperatures ($5.1\leq T\leq 10$~K), which display no signs of magnetic order, as evidenced buy the lack of hyperfine splitting. This does not completely rule out the possibility of exhibiting magnetic order below the lowest measured temperature, but it is unlikely as argued below. 

\begin{figure}[H]
 \centering
 \includegraphics[width=\linewidth]{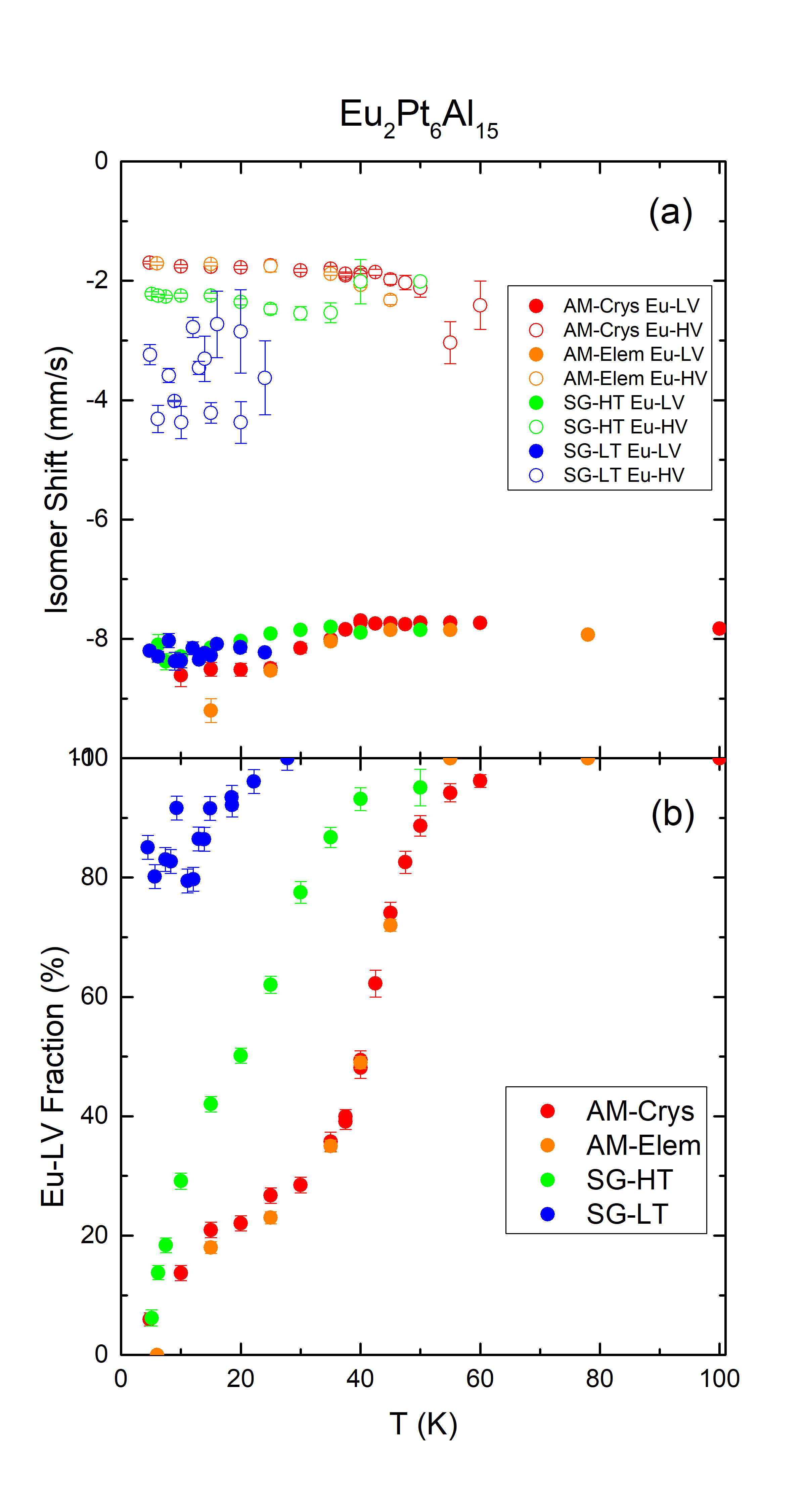}
 \caption{{(a) Isomer shifts corresponding to Eu-LV (solid symbols) and Eu-HV (open symbols), for the arc-melted crystals (red), the arc-melted elements (orange), the solution grown crystals decanted at highest temperatures (green), and those decanted at lower temperatures (blue). (b) temperature dependence of the Eu-LV fraction for the arc-melted crystals (red), the arc-melted elements (orange), the solution grown crystals decanted at highest temperatures (green), and those decanted at lower temperatures (blue).}}
 \label{fig:Mossbauer_intermediate}
\end{figure}

All of the spectra measured can be adequately fitted by two contributions with different isomer shifts, with varying relative weights. Figure \ref{fig:Mossbauer_intermediate} summarizes the parameters obtained from each M\"ossbauer spectrum as a function of temperature. In particular, Fig. \ref{fig:Mossbauer_intermediate}(a) shows the isomer shifts corresponding to Eu-LV (solid symbols) and Eu-HV (open symbols), for the arc-melted crystals (red), the arc-melted elements (orange), the solution grown crystals decanted at highest temperatures (green), and those decanted at lower temperatures (blue). These results show no marked difference between the isomer shifts of the Eu-LV ions in each phase, but do reflect some difference in the isomer shift of the Eu-HV, yielding an intermediate value for the solution-grown crystals decanted at higher temperature (SG-HT). Figure \ref{fig:Mossbauer_intermediate}(b) shows the temperature dependence of the Eu-LV fraction, demonstrating that the Eu ions in the SG-HT samples fully transform into the higher valence at the lowest temperatures and that this transformation occurs at lower temperatures than for the AM samples. Given that 94\% of the Eu ions of the SG-HT phase are in the HV state (close to the non-moment baring Eu$^{3+}$) at 5.1~K, it is unlikely that localized magnetic order emerges below this temperature. Single-crystal XRD measurements would be needed in order to compare the structure of this phase with the other two, to further understand which specific structural parameters are tuning the valence transition in this system.

\newpage
\nocite{apsrev41Control}
\bibliographystyle{apsrev4-1}
\bibliography{EuPtAl.bib}

\end{document}